\documentclass[aps,prb,twocolumn,superscriptaddress,floatfix]{revtex4-2}
\usepackage{graphicx}
\usepackage{amsmath,amssymb,amsfonts,bm}
\usepackage{dcolumn}
\usepackage{xcolor}
\usepackage{txfonts}
\usepackage{physics}
\usepackage{hyperref}
\usepackage{tikz}
\usetikzlibrary{arrows.meta,decorations.pathreplacing}

\hypersetup{hidelinks}

\begin{document}

\title{Vortex-core Majorana coupling to a chiral edge in a \(p_x+ip_y\) superconductor: Nonmonotonic spectral reorganization and coherent fermion-parity dynamics}
\author{Peiyao Liu}
\affiliation{Institute of Physics, Chinese Academy of Sciences, Beijing 100190, China}
\affiliation{School of Physical Sciences, University of Chinese Academy of Sciences, Beijing 101408, China}
\author{Yi Zhou}
\affiliation{Institute of Physics, Chinese Academy of Sciences, Beijing 100190, China}
\date{\today}

\begin{abstract}
We study how vortex--edge coupling reorganizes the low-energy sector of a finite two-dimensional \(p_x+ip_y\) superconducting disk as a function of the vortex--boundary separation \(d\) and examine what this reorganization implies for the parity memory associated with a prescribed vortex-core Majorana wave packet, a resource relevant to Majorana-based quantum operations. Bogoliubov--de Gennes calculations reveal nonmonotonic core--edge reorganization of the lowest positive-energy finite-disk eigenstate, with particularly rapid variation near \(d\simeq7\xi\), where \(\xi\) is the coherence length. To separate this eigenstate reorganization from the spectral representation of a prescribed state, we rigidly translate a centered-vortex core-reference packet to each fixed vortex position, restrict it to the target disk, and project it, without intermediate normalization, onto the particle-hole-complete low-energy subspace. For \(\Delta_0/E_F=0.36\) and disk radius \(R=30\xi\), the resulting retained norm exceeds \(0.98\) at all six sampled separations, \(4.25\leq d/\xi\leq8.25\), while, depending on \(d\), the spectral measure is concentrated near zero energy, fragmented over several low-energy levels, or dominated by finite-energy weight. Correspondingly, the signed parity correlator displays slow temporal variation, rapid coherent dephasing, or sign-changing oscillations, with possible finite-size recurrences at later times. Thus a large retained norm does not by itself imply spectral concentration or persistent parity memory.
\end{abstract}

\maketitle

\section{Introduction}\label{sec:introduction}

Two-dimensional chiral topological superconductors provide a canonical setting in which Majorana degrees of freedom occur at both defects and boundaries. In the weak-pairing phase of a spinless \(p_x+ip_y\) superconductor, an isolated unit-winding vortex binds a Majorana zero mode, while an open boundary supports a chiral Majorana mode~\cite{Read_2000,Qi_2011,Sato_2017,Stone_2004,Ivanov_2001}. The observation of the even-denominator \(\nu=5/2\) fractional quantum Hall state motivated several paired-state descriptions with non-Abelian topological order~\cite{Willett_1987,Willett_2013,Wen_1991,Moore_Read_1991,Blok_Wen_1992}. In particular, the Moore--Read state supports non-Abelian quasiholes and shares the topological structure of the weak-pairing superconductor~\cite{Read_Rezayi_1996,Read_2000,Nayak_2008}. The coexistence of vortex and edge Majoranas therefore raises a basic finite-geometry question: how does their coupling reorganize the low-energy sector as a function of the vortex--boundary separation?

At large vortex--boundary separation, the core and edge modes are approximately spatially distinct. At smaller separations, overlap of their wave-function tails allows the vortex-core mode to hybridize with states of the low-energy chiral edge branch, which is gapless in the large-system limit. This situation contrasts with the case of a vortex near the boundary of a conventional fully gapped \(s\)-wave superconductor, where the boundary modifies the Caroli--de Gennes--Matricon (CdGM) spectrum but provides no gapless edge channel~\cite{Caroli_1964,Kopnin_1991,Mel_nikov_2009}. For a centered vortex, rotational symmetry conserves angular momentum and restricts each core state to symmetry-compatible edge channels~\cite{Kraus_2009}; an off-center vortex relaxes this selection rule and permits coupling to multiple edge channels. The resulting coupling can therefore reorganize the low-energy sector broadly rather than produce a simple monotonic splitting of two isolated modes.

This problem involves two distinct physical questions. The first concerns how individual finite-disk eigenstates redistribute between the vortex core and the boundary. The second concerns whether a prescribed core-localized self-conjugate packet remains spectrally concentrated and, when paired with an ideal stationary remote Majorana, retains the parity memory of the resulting fermionic mode. Spatial localization alone does not resolve the second question: a compact packet can span several eigenfrequencies whose components dephase under unitary evolution. Their destructive interference in \(C(t)\), termed coherent dephasing here, can produce sign changes and finite-size recurrences~\cite{Gogolin_2016,Rigol_2008}. This distinction matters for Majorana-based operations, where overlap and dynamical hybridization can generate coherent errors~\cite{Cheng_2009,Hodge_2025}, while environmental fluctuations introduce additional qubit dephasing~\cite{Knapp_2018}. In our finite-disk calculations, the vortex is fixed at each separation; the translated-packet construction is an idealized reference preparation, not a model of physical vortex motion.

A phenomenological low-energy description of this coupling was developed by Fendley, Fisher, and Nayak using boundary conformal field theory~\cite{Fendley_2009}. Starting from the neutral edge sector of a Moore--Read quantum Hall droplet, which is formally equivalent to the edge of a \(p_x+ip_y\) superconductor, they ``squashed'' the chiral system into an effectively one-dimensional strip with left- and right-moving Majorana fields and coupled a localized vortex-core Majorana zero mode to those fields. This resonant coupling generates a crossover scale proportional to the squared tunneling amplitude and drives a boundary-condition flow corresponding to hybridization of the localized zero mode with the edge. Combined with the expected exponential suppression of tunneling at large vortex--boundary separation, the theory yields a smooth crossover envelope. As a universal low-energy description, however, it does not resolve the microscopic oscillatory wave-function tails and CdGM states of a Bogoliubov--de Gennes (BdG) system or the associated multilevel spectral reorganization at vortex--boundary separations of only a few coherence lengths.

We find that finite-disk eigenstates exhibit strongly nonmonotonic core--edge reorganization as a function of the vortex--boundary separation \(d\), with particularly rapid variation near \(d\simeq7\xi\), where \(\xi\) is the coherence length. For \(\Delta_0/E_F=0.36\), the retained norm of the translated core-reference packet after disk restriction and low-energy projection exceeds \(0.98\) at all six sampled separations, while, depending on \(d\), its spectral measure is concentrated near zero energy, fragmented over several low-energy levels, or dominated by finite-energy weight. Correspondingly, the parity dynamics display slow temporal variation, rapid coherent dephasing, or sign-changing oscillations. Thus a large retained norm does not guarantee concentration in a single eigenmode or persistent parity memory.

The remainder of the paper is organized as follows. Section~\ref{sec:model} introduces the model and numerical framework; Sec.~\ref{sec:spectrum} analyzes core--edge hybridization in finite-disk eigenstates; Sec.~\ref{sec:parity_decay} constructs the core-reference packet and examines its spectral measure and parity dynamics; and Sec.~\ref{sec:conclusion} discusses the physical implications and limitations. The appendices provide numerical validation and supporting derivations.

\section{Finite-disk model and numerical framework}\label{sec:model}

We formulate the finite-disk BdG problem used throughout this work. We first specify the continuum Hamiltonian and particle-hole convention, then define the prescribed vortex profile and boundary conditions on the disk, and finally introduce the dimensionless parameters and numerical implementation.

\subsection{Continuum BdG model and particle-hole symmetry}

We consider a two-dimensional spinless \(p_x+ip_y\) superconductor with a spatially varying complex pairing field \(\Delta(\mathbf r)\). Its grand-canonical mean-field Hamiltonian, \(\mathcal H_{\rm MF}\equiv H-\mu\hat N\), is~\cite{Read_2000}
\begin{equation}
    \begin{aligned}
        \mathcal H_{\rm MF}
        =\int\! \mathrm d^2r\,\Bigg\{&
        c_{\mathbf r}^\dagger
        \left(-\frac{\hbar^2}{2m_e}\nabla^2-\mu\right)c_{\mathbf r}\\
        &+\frac{1}{4k_F}\left[
        c_{\mathbf r}^\dagger
        \{\Delta(\mathbf r),\partial_x+i\partial_y\}
        c_{\mathbf r}^\dagger+\mathrm{H.c.}\right]\Bigg\}.
    \end{aligned}
    \label{eq:bdg_hamiltonian}
\end{equation}
Here \(c_{\mathbf r}\) annihilates a fermion at \(\mathbf r\), \(m_e\) is the effective mass, \(\mu\) is the chemical potential, and \(\hat N\) is the particle-number operator. The anticommutator provides the spatial symmetrization of the inhomogeneous \(p\)-wave pairing operator required for a Hermitian BdG operator under the imposed boundary conditions: acting on a quasiparticle amplitude, it differentiates both the amplitude and the gap profile. We take \(\mu>0\), placing the model in the topological weak-pairing phase, and let \(\Delta_0>0\) denote the asymptotic pairing amplitude. We identify \(E_F=\mu\) and define \(k_F=\sqrt{2m_e\mu}/\hbar\) and \(v_F=\hbar k_F/m_e\). When the additional BCS weak-coupling condition \(\Delta_0/\mu\ll1\) holds, the minimum bulk excitation gap approaches \(\Delta_0\)~\cite{Read_2000}.

For a quasiparticle creation operator \(\psi^\dagger=\int\! \mathrm d^2r\,[u(\mathbf r)c_{\mathbf r}^\dagger+v(\mathbf r)c_{\mathbf r}]\), the equation of motion \([\mathcal H_{\rm MF},\psi^\dagger]=E\psi^\dagger\) yields the BdG equations~\cite{DeGennes}
\begin{align}
    \left(-\frac{\hbar^2}{2m_e}\nabla^2-\mu\right)u(\mathbf r)
    +\frac{1}{2k_F}
    \{\Delta(\mathbf r),\partial_x+i\partial_y\}v(\mathbf r)
    &=Eu(\mathbf r),\nonumber\\
    -\frac{1}{2k_F}
    \{\Delta^*(\mathbf r),\partial_x-i\partial_y\}u(\mathbf r)
    -\left(-\frac{\hbar^2}{2m_e}\nabla^2-\mu\right)v(\mathbf r)
    &=Ev(\mathbf r).
    \label{BdG}
\end{align}
With the Nambu spinor \(\Psi=(u,v)^\top\), these equations read \(H_{\rm BdG}\Psi=E\Psi\). The BdG Hamiltonian obeys \(\mathcal C H_{\rm BdG}\mathcal C^{-1}=-H_{\rm BdG}\), where \(\mathcal C=\tau_xK\), \(\tau_x\) acts in Nambu space, and \(K\) denotes complex conjugation. Consequently, if \(\Psi_E=(u,v)^\top\) has energy \(E\), then \(\mathcal C\Psi_E=(v^*,u^*)^\top\) has energy \(-E\). Nonzero-energy states therefore occur in particle-hole pairs, and we take the \(E>0\) member of each pair as the independent fermionic mode.

\begin{figure}[tb]
    \centering
    \begin{tikzpicture}[x=1cm,y=1cm,line cap=round,line join=round]
        \path[use as bounding box] (0,0) rectangle (7.2,4.5);
        \fill[black!4] (0.35,0.55) rectangle (6.85,3.20);
        \draw[line width=0.9pt] (0.35,3.20) -- (6.85,3.20);
        \node[anchor=west] at (0.55,3.62) {vacuum};
        \node[anchor=west] at (0.55,2.78) {superconductor};

        \coordinate (V) at (3.55,1.18);
        \filldraw[fill=red!78!black,draw=black,line width=0.65pt]
            (V) circle[radius=3.2pt];
        \node[anchor=east] at (3.28,1.18) {vortex};

        \draw[densely dotted,line width=0.55pt] (3.72,1.18) -- (5.05,1.18);
        \draw[
            {Stealth[length=5pt,width=4pt]}-{Stealth[length=5pt,width=4pt]},
            line width=0.75pt
        ] (5.05,1.18) -- (5.05,3.20);
        \node[anchor=west] at (5.18,2.19) {$d$};
        \node at (3.60,0.16) {(a)};
    \end{tikzpicture}

    \begin{tikzpicture}[x=1.2cm,y=1.2cm,line cap=round,line join=round]
        \path[use as bounding box] (-2.05,-2.48) rectangle (3.10,2.02);
        \coordinate (O) at (0,0);
        \coordinate (P) at (0.35,1.30);
        \coordinate (A) at (1.15,0);
        \coordinate (B) at (2.226,0);

        \fill[black!4] (O) circle[radius=1.60];
        \draw[line width=0.75pt] (O) circle[radius=1.60];
        \draw[line width=0.65pt] (O) -- (B);
        \draw[line width=0.75pt,-{Stealth[length=5pt,width=4pt]}] (O) -- (P);
        \draw[line width=0.65pt] (P) -- (A);
        \draw[line width=0.65pt] (P) -- (B);

        \draw[line width=0.55pt] (P) ++(254.9:0.34)
            arc[start angle=254.9,end angle=301.6,radius=0.34];
        \draw[line width=0.55pt] (P) ++(254.9:0.57)
            arc[start angle=254.9,end angle=325.3,radius=0.57];
        \draw[line width=0.55pt] (A) ++(121.6:0.27)
            arc[start angle=121.6,end angle=180,radius=0.27];

        \fill (O) circle[radius=1.1pt];
        \filldraw[fill=red!78!black,draw=black,line width=0.65pt]
            (A) circle[radius=3.2pt];
        \filldraw[fill=blue!70!black,draw=black,line width=0.65pt]
            (B) circle[radius=3.2pt];

        \node[anchor=north east] at (-0.06,-0.05) {$O$};
        \node[anchor=south east] at (0.10,0.71) {$\mathbf r$};
        \node[anchor=south] at (0.57,-0.34) {$r_1$};
        \node[anchor=north] at (1.18,-0.05) {$\mathbf r_1$};
        \node[anchor=north] at (2.256,-0.05) {$\mathbf r_2$};
        \node at (0.39,0.85) {\footnotesize$\alpha_1$};
        \node at (0.6,0.62) {\footnotesize$\alpha_2$};
        \node at (0.78,0.22) {\footnotesize$\alpha_2$};

        \draw[line width=0.55pt] (0,-1.78) -- (2.226,-1.78);
        \draw[line width=0.55pt] (0,-1.69) -- (0,-1.87);
        \draw[line width=0.55pt] (2.226,-1.69) -- (2.226,-1.87);
        \node[anchor=north] at (1.113,-1.82) {$R^2/r_1$};
        \node at (0.55,-2.34) {(b)};
    \end{tikzpicture}
    \caption{Vortex geometry and image construction. (a) Local view of a vortex at separation \(d\) from the boundary between the two-dimensional \(p_x+ip_y\) superconductor and vacuum. (b) Disk of radius \(R\) containing the physical vortex (red) at \(\mathbf r_1=(r_1,0)\) and the image antivortex (blue) at \(\mathbf r_2=(R^2/r_1,0)\) outside the disk. The auxiliary angles \(\alpha_1\) and \(\alpha_2\) specify the geometry used to impose the phase boundary condition. The resulting image phase enforces \(\partial_r\phi|_{r=R}=0\) for the order-parameter phase \(\phi\), eliminating the normal phase-gradient contribution to the condensate current. The vortex--boundary separation is \(d=R-r_1\).}
    \label{fig:setup}
    \label{fig:disk_image}
\end{figure}
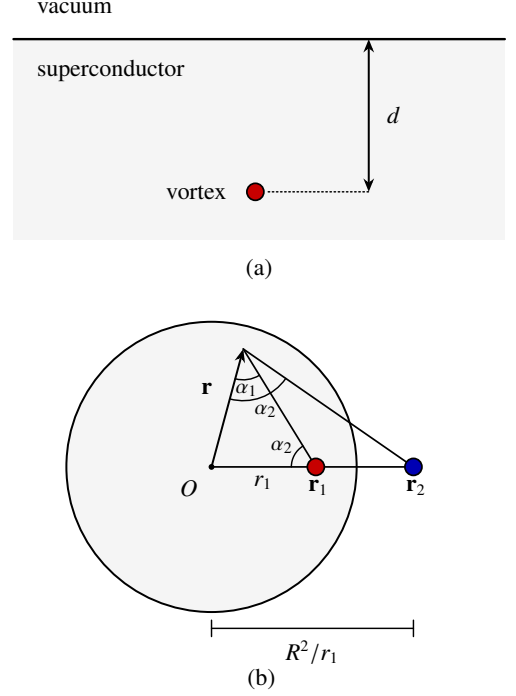

\subsection{Disk geometry, boundary conditions, and vortex profile}

An isolated unit-winding vortex can be represented as~\cite{Abrikosov_1957,Abrikosov_2004}
\begin{equation*}
    \Delta(r,\theta)=\Delta_0 f(r/\xi)e^{i\theta},
    \qquad f(0)=0,\quad f(\infty)=1.
\end{equation*}
Such a vortex supports a Caroli--de Gennes--Matricon (CdGM) ladder with characteristic level spacing of order \(\Delta_0^2/E_F\)~\cite{Caroli_1964,Kopnin_1991}. To study its coupling to a single smooth boundary while avoiding a periodic vortex array and the corners of a finite square~\cite{ZhangZhaZhou2019}, we place the vortex inside a disk of radius \(R\), as shown in Fig.~\ref{fig:setup}. At \(r=R\), the quasiparticle amplitudes obey the hard-wall condition \(u=v=0\), while the condensate phase is independently constrained by \(\partial_r\phi|_{r=R}=0\).

We use a prescribed order-parameter profile. Let the physical vortex be at \(\mathbf r_v\equiv\mathbf r_1=(r_1,0)\) and introduce an image antivortex at \(\mathbf r_2=(R^2/r_1,0)\) outside the disk. Using the method of images for the phase~\cite{Sobnack_2001} and a product ansatz for the amplitude, we take
\begin{equation}
    \begin{aligned}
        \Delta(\mathbf r)
        &=\Delta_0
        f\left(\frac{|\mathbf r-\mathbf r_1|}{\xi}\right)
        f\left(\frac{|\mathbf r-\mathbf r_2|}{\xi}\right)\\
        &\quad\times\exp\!\left\{i\left[
        \theta(\mathbf r-\mathbf r_1)
        -\theta(\mathbf r-\mathbf r_2)+\pi\right]\right\},
    \end{aligned}
    \label{eq:disk_image}
\end{equation}
Here \(f(s)=\tanh s\) sets the vortex-core profile~\cite{Tinkham_2004}, and the vortex--boundary separation is \(d=R-r_1\). Writing \(\Delta=|\Delta|e^{i\phi}\), the phase-gradient contribution to the normal condensate current is proportional to \(|\Delta|^2\partial_r\phi\)~\cite{Tinkham_2004}. Direct differentiation using the geometry in Fig.~\ref{fig:disk_image} gives \(\partial_r\phi|_{r=R}=0\). The image phase therefore eliminates the normal phase-gradient contribution to the boundary current, while \(u=v=0\) separately imposes the quasiparticle hard wall. The image-amplitude factor is nearly unity at the six separations used for the translated-packet analysis; Appendix~\ref{app:image_amplitude_sensitivity} shows that setting it to unity at \(d=2\xi\) leaves the principal spectral and core-weight features unchanged.

We neglect the vector potential and work in the extreme-type-II limit, retaining the prescribed phase texture while omitting magnetic screening and orbital coupling~\cite{DeGennes,Tinkham_2004}. This finite-disk approximation assumes a London penetration depth \(\lambda\gg R=30\xi\).

In the limit \(r_1\to0^+\), the image position satisfies \(|\mathbf r_2|\to\infty\), and Eq.~\eqref{eq:disk_image} reduces to \(\Delta_0f(r/\xi)e^{i\theta}\). We use this limiting centered-vortex profile to define the core-reference packet in Sec.~\ref{subsec:bdg_survival}, where the lowest particle-hole pair supplies the subspace from which the packet is constructed. Writing the pairing components as \(\Delta_x\) and \(\Delta_y\), the ansatz fixes \(\Delta_y/\Delta_x=i\) throughout the disk, defining the fixed-chirality model used below.

\subsection{Dimensionless formulation and numerical implementation}

Using the coherence length \(\xi=\hbar v_F/(\pi\Delta_0)\), we express lengths in units of \(\xi\) and energies in units of \(\Delta_0\)~\cite{Tinkham_2004}. We define \(\mathbf l=\mathbf r/\xi\), \(l=|\mathbf l|\), \(U(\mathbf l)=u(\mathbf r)\), \(V(\mathbf l)=v(\mathbf r)\), and \(\bar\Delta(\mathbf l)=\Delta(\mathbf r)/\Delta_0\). Dividing Eq.~\eqref{BdG} by \(\Delta_0\) then gives
\begin{align}
    \left(-\frac{\pi^2\Delta_0}{4\mu}\nabla_{\mathbf l}^2
    -\frac{\mu}{\Delta_0}\right)U
    +\frac{\pi\Delta_0}{4\mu}
    \{\bar\Delta,\partial_{l_x}+i\partial_{l_y}\}V
    &=\frac{E}{\Delta_0}U,\nonumber\\
    -\frac{\pi\Delta_0}{4\mu}
    \{\bar\Delta^*,\partial_{l_x}-i\partial_{l_y}\}U
    -\left(-\frac{\pi^2\Delta_0}{4\mu}\nabla_{\mathbf l}^2
    -\frac{\mu}{\Delta_0}\right)V
    &=\frac{E}{\Delta_0}V.
    \label{eq:dimensionless_bdg}
\end{align}
For a fixed dimensionless gap profile, the material dependence enters through the single ratio \(\Delta_0/E_F\), while the dimensionless geometry is specified by the disk radius \(L=R/\xi\) and vortex position \(\mathbf l_1=\mathbf r_1/\xi\), with \(d/\xi=L-|\mathbf l_1|\). The scaled gap profile follows directly from Eq.~\eqref{eq:disk_image}.

We set $L=30$ and study $\Delta_0/E_F=0.08$ and $0.36$, spanning a weak-coupling regime and a regime motivated by iron-based vortex systems. Zero-bias vortex states consistent with Majorana modes have been reported on the surface of $\mathrm{FeTe}_{0.55}\mathrm{Se}_{0.45}$~\cite{Zhang_2018}; effective Dirac-cone calculations have used $\Delta_0/E_F=0.36$~\cite{Chiu_2020}; and CdGM measurements in FeSe-based systems yield ratios from approximately $0.18$ to values of order unity~\cite{Chen_2020,Chen_2018}. The cited results motivate the parameter range used in the present spinless parabolic-band model.

We discretize Eq.~\eqref{eq:dimensionless_bdg} on a polar grid containing $N_l$ radial points, $N_\theta$ angular points, and one additional point at the origin. The angular direction is periodic, and the hard-wall condition $U=V=0$ is imposed at $l=L$. We use $(N_l,N_\theta)=(900,900)$ for $\Delta_0/E_F=0.08$ and $(500,500)$ for $\Delta_0/E_F=0.36$. The weaker-coupling case requires the finer grid because its larger \(k_F\xi=2E_F/(\pi\Delta_0)\) produces a shorter Fermi wavelength in coherence-length units. Appendix~\ref{sec:fd_scheme} describes the finite-difference construction and its weighted self-adjoint form.

For each geometry, we retain the $N=50$ lowest positive-energy eigenstates for the subsequent low-energy analysis. In the centered-vortex limit, this set includes the near-zero fermionic state formed from the vortex and edge Majorana components, as well as the CdGM angular-momentum channels $m=1$ and $2$.

\section{Core--edge hybridization in finite-disk eigenstates}\label{sec:spectrum}

Here we characterize static core--edge hybridization in individual positive-energy finite-disk eigenstates as a function of \(d\). We first introduce regional diagnostics of complementary Majorana separation, then use the centered vortex as a reference, and finally compare the nonmonotonic spectral reorganization with tail-overlap and interference estimates. Section~\ref{sec:parity_decay} separately analyzes the translated self-conjugate packet and its parity dynamics.

\begin{figure*}[tb]
    \centering
    \includegraphics[width=\textwidth]{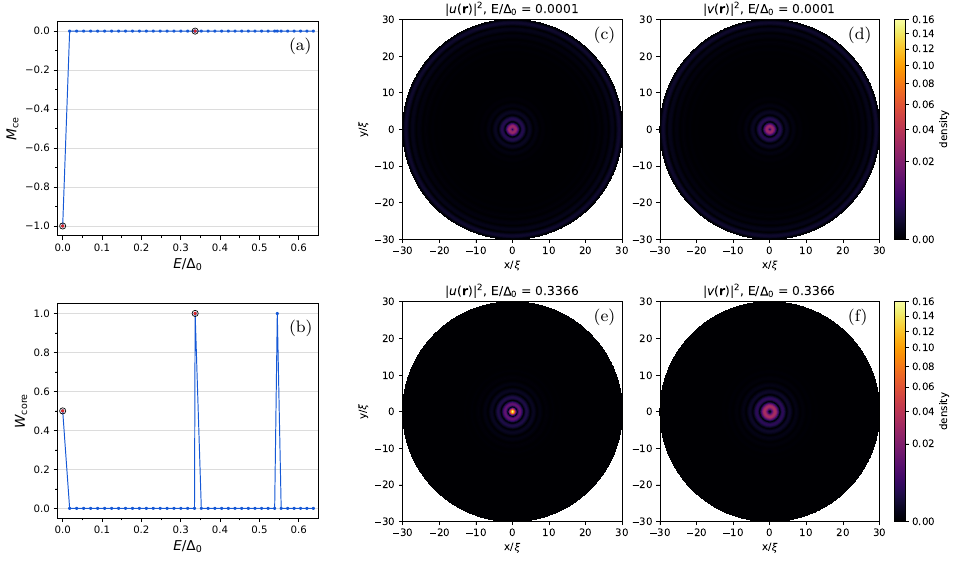}
    \caption{Majorana-polarization and regional-weight diagnostics for a centered vortex at $\Delta_0/E_F=0.36$, $R=30\xi$, and $d=30\xi$. (a) Core--edge Majorana polarization $M_{\rm ce}$ and (b) core weight $W_{\rm core}$ of the low-lying positive-energy eigenstates. The circled red dots mark the two states whose densities are shown in panels (c)--(f). The lowest state has $E_0=1.44\times10^{-4}\Delta_0$, $M_{\rm ce}=-0.9997$, and $W_{\rm core}=0.5000$, indicating oppositely polarized, equally weighted core and edge Majorana components. (c) Particle density $|u|^2$ and (d) hole density $|v|^2$ of this state. For comparison, (e) $|u|^2$ and (f) $|v|^2$ of the core-localized CdGM state at $E=0.3366\Delta_0$, which has $W_{\rm core}=0.9998$ but $\lvert M_{\rm ce}\rvert<10^{-9}$. Thus strong core localization alone does not identify separated Majorana components.}
    \label{fig:fd_polar_d30_cdgm}
\end{figure*}

\begin{figure*}[tb]
    \centering
    \includegraphics[width=\textwidth]{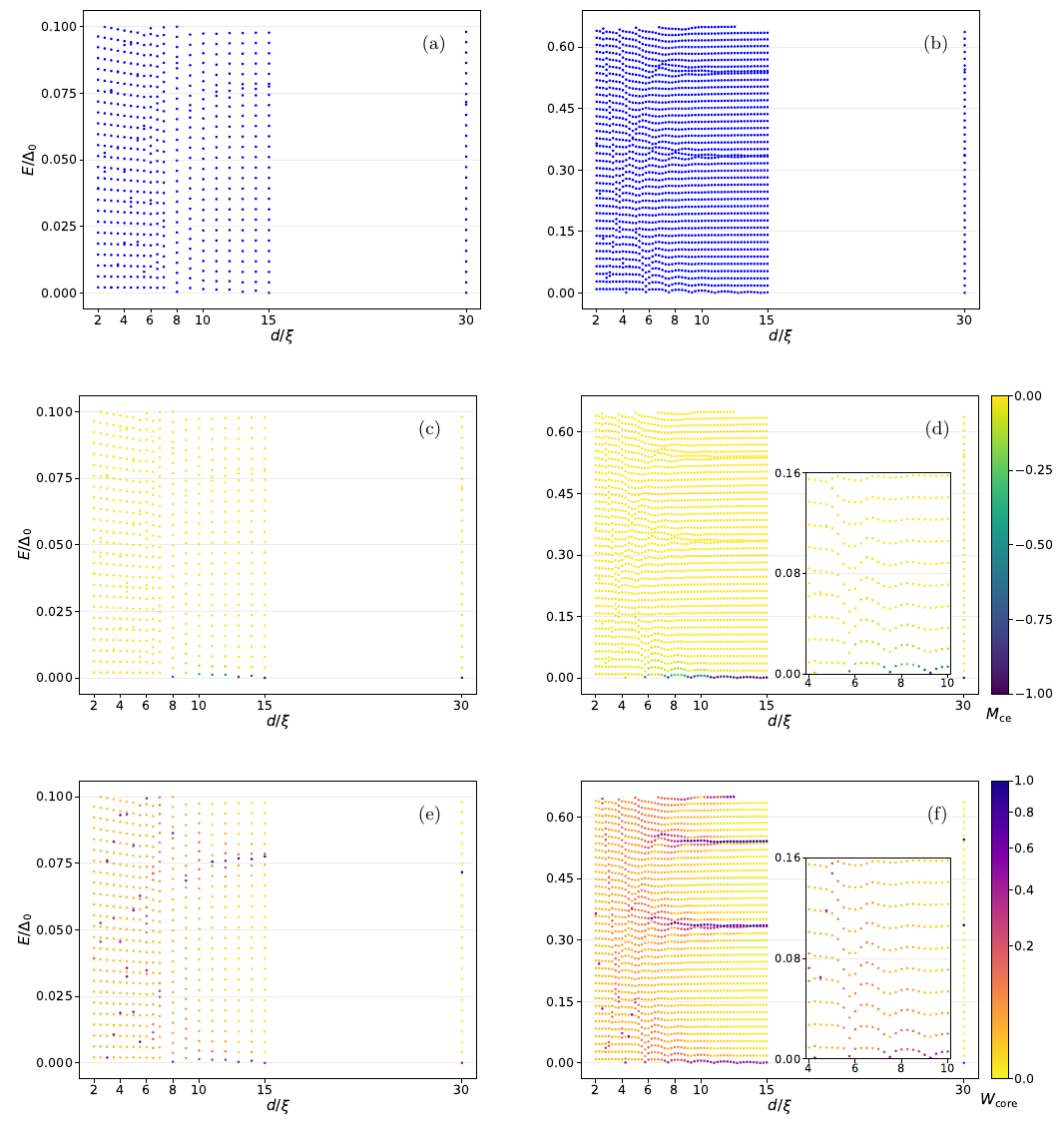}
    \caption{Positive-energy spectrum and regional diagnostics versus vortex--boundary distance for $L=R/\xi=30$. The columns correspond to $\Delta_0/E_F=0.08$ and $0.36$. (a) and (b) Eigenenergies. (c) and (d) The same levels colored by the core--edge Majorana polarization $M_{\rm ce}$, with dark blue indicating $M_{\rm ce}\simeq-1$. (e) and (f) Levels colored by the regional core weight $W_{\rm core}$, with dark blue indicating $W_{\rm core}\simeq1$. The insets in panels (d) and (f) enlarge the low-energy region \(0\leq E/\Delta_0\leq0.16\) for \(4\leq d/\xi\leq10\). The diagnostics are evaluated for each unrecombined positive-energy eigenstate using the distance-dependent partition in Eq.~\eqref{eq:core_edge_partition}; as $\Omega_{\rm core}$ contracts with decreasing $d$, $W_{\rm core}$ tracks regional probability redistribution. For $\Delta_0/E_F=0.36$, the coupling is already appreciable at larger $d$, with particularly rapid nonmonotonic reorganization near $d\simeq7\xi$.}
    \label{fig:energy_vs_d_observables}
\end{figure*}

\subsection{Regional diagnostics of core--edge Majorana separation}

Every nonzero-energy BdG eigenstate and its particle-hole partner span a two-dimensional subspace that admits a self-conjugate basis. Self-conjugacy alone therefore does not diagnose spatially separated Majorana components. We instead ask whether the particle-hole pair associated with an individual positive-energy eigenstate can be recombined into components supported predominantly near the vortex and the edge. Regional Majorana polarization probes this separation, whereas regional probability weight measures only spatial support~\cite{Sticlet_2012_MP,Sedlmayr_2015_MP,Bena_2017_MP,Awoga_2024_MP}.

We partition the disk \(D=\{\mathbf r:|\mathbf r|<R\}\) by assigning each point to the region closer to the vortex or the boundary:
\begin{equation}
    \begin{aligned}
        \Omega_{\rm core}
        &=\{\mathbf r\in D:|\mathbf r-\mathbf r_v|<R-|\mathbf r|\},\\
        \Omega_{\rm edge}&=D\setminus\Omega_{\rm core}.
    \end{aligned}
    \label{eq:core_edge_partition}
\end{equation}
Along the radial line from the vortex to the nearest boundary point, the interface lies a distance \(d/2\) from the vortex. Consequently, \(\Omega_{\rm core}\) contracts with decreasing \(d\) and defines the core side of this distance-adapted diagnostic partition. For a normalized positive-energy eigenstate \(\psi=(u,v)^\top\) and a region \(A\in\{\Omega_{\rm core},\Omega_{\rm edge}\}\) with \(W_A>0\), we define
\begin{equation}
    W_A=\int_A\! \mathrm d^2r\,(|u|^2+|v|^2),
    \qquad
    P_A=\frac{2\int_A\! \mathrm d^2r\,(uv)^*}{W_A}.
    \label{eq:regional_weight}
\end{equation}
Here \(W_A\) is the probability weight in region \(A\), while \(P_A\) is its normalized regional particle--hole coherence. The phase dependence of \(P_A\) cancels in the relative polarization
\begin{equation}
    M_{\rm ce}
    =\operatorname{Re} \left(P_{\Omega_{\rm core}}P_{\Omega_{\rm edge}}^*\right),
    \qquad -1\leq M_{\rm ce}\leq0.
    \label{eq:core_edge_majorana_polarization}
\end{equation}
so \(M_{\rm ce}\) is invariant under a global phase change of the BdG eigenstate. For a nonzero-energy state and complementary regions covering the disk, orthogonality of the particle-hole partners gives \(-1\leq M_{\rm ce}\leq0\), as derived in Appendix~\ref{app:majorana_polarization}. A value \(M_{\rm ce}\simeq-1\) indicates complementary self-conjugate components supported predominantly in the core and edge regions, whereas \(M_{\rm ce}\simeq0\) indicates that this partition resolves no such separation; the latter includes components concentrated in the same region and generic extended states.

\(W_{\rm core}\equiv W_{\Omega_{\rm core}}\) measures the probability weight on the vortex side of the distance-adapted partition and therefore tracks regional redistribution as \(d\) varies. We evaluate both \(W_{\rm core}\) and \(M_{\rm ce}\) for each unrecombined positive-energy eigenstate; Appendix~\ref{app:majorana_polarization} examines their sensitivity to the regional partition.

\subsection{Centered-vortex reference and distance dependence}

We use the centered vortex as a symmetry-controlled reference. At \(d=R=30\xi\), the lowest positive-energy state forms a near-zero fermionic mode composed of spatially separated vortex and edge Majorana components, whereas higher CdGM states are ordinary core-localized excitations.

Figure~\ref{fig:fd_polar_d30_cdgm} illustrates this distinction. The lowest centered-vortex state has \(E_0=1.44\times10^{-4}\Delta_0\), \(M_{\rm ce}=-0.9997\), and \(W_{\rm core}=0.5000\), indicating oppositely polarized Majorana components of equal weight in the core and edge regions. The CdGM excitation at \(E=0.3366\Delta_0\), by contrast, has \(W_{\rm core}=0.9998\) and \(M_{\rm ce}\simeq0\): it is strongly core localized without complementary core--edge polarization. Its energy differs by only \(0.38\%\) from the independent centered-vortex benchmark, \(0.3379\Delta_0\), in Appendix~\ref{sec:centered_vortex}. Thus, \(W_{\rm core}\) alone does not diagnose Majorana separation.

Figure~\ref{fig:energy_vs_d_observables} extends the analysis to off-center vortices for both values of \(\Delta_0/E_F\). The \(\Delta_0/E_F=0.08\) results provide a weak-coupling comparison, while the denser distance sampling at \(\Delta_0/E_F=0.36\) supports the detailed analysis below. At large \(d\), the lowest positive-energy state has \(M_{\rm ce}\simeq-1\) and \(W_{\rm core}\simeq1/2\), consistent with separated core and edge Majorana components. For \(\Delta_0/E_F=0.36\), the lowest-state polarization takes the values \(-0.4612\), \(-0.8159\), and \(-0.3882\) at \(d=10\xi\), \(7.5\xi\), and \(7.25\xi\), respectively, displaying pronounced nonmonotonic variation. Between the latter two separations, \(W_{\rm core}\) changes from \(0.4657\) to \(0.2904\). The corresponding densities are shown in Appendix~\ref{app:majorana_polarization}.

The pronounced nonmonotonic variation identifies \(d\simeq7\xi\) as the region of strongest reorganization among the sampled separations. At \(\Delta_0/E_F=0.08\) and \(d=5\xi\), the qualitative \(W_{\rm core}(E)\) pattern appears on both grids in the simultaneous radial-and-angular comparison of Appendix~\ref{sec:fd_scheme}. Higher CdGM states can remain strongly core weighted while exhibiting little complementary Majorana separation.

\subsection{Tail-overlap and interference estimates}

The asymptotic vortex and edge tails provide two distance scales for interpreting the numerical behavior: a broad overlap scale set by exponential decay and a shorter modulation scale set by oscillatory phases. We use these scales as order-of-magnitude guides to the observed spectral reorganization.

Outside the vortex core, the zero-energy bulk dispersion has complex wave numbers. For $0<\Delta_0/\mu<2$, their real part is
\begin{equation}
    k_{\rm osc}=k_F\sqrt{1-\frac{1}{4}\left(\frac{\Delta_0}{\mu}\right)^2}.
\end{equation}
At zero tangential momentum, the complex bulk wave number is \(k_{\rm osc}+i/(\pi\xi)\), so both the vortex tail and the normal tail in the \(q=0\) edge-mode limit decay over \(\pi\xi\). Let \(x=R-r\geq0\) denote the inward normal distance and take the normal dependence to be \(e^{ik_\perp x}\). For a locally flat edge mode with tangential momentum \(q\), the two decaying normal-momentum branches are
\[
    k_\perp(q)
    =\pm\sqrt{k_{\rm osc}^2-q^2}+\frac{i}{\pi\xi}.
\]
For the low-lying disk channels, \(q=j/R\), where \(j\) denotes the dimensionless angular-momentum label, and, at fixed \(j\),
\[
    \begin{aligned}
        \operatorname{Re}k_\perp
        &=\pm k_{\rm osc}\left[
        1-\frac{j^2}{2(k_{\rm osc}R)^2}
        +\mathcal O\bigl((k_{\rm osc}R)^{-4}\bigr)
        \right],\\
        \operatorname{Im}k_\perp
        &=\frac{1}{\pi\xi},
    \end{aligned}
\]
within the locally flat approximation.

For the centered profile \(\Delta(\mathbf r)=\Delta_0\tanh(r/\xi)e^{i\theta}\), the weak-coupling Andreev approximation of Cheng \emph{et al.} gives the isolated-vortex zero-mode profile~\cite{Cheng_2009}
\begin{equation}
    \binom{U(\mathbf l)}{V(\mathbf l)}
    \simeq
    \mathcal N_v(\cosh l)^{-1/\pi}
    \binom{e^{i\theta}J_1(k_F\xi l)}
          {e^{-i\theta}J_1(k_F\xi l)} ,
    \label{eq:approx_cdgm}
\end{equation}
where \(\mathcal N_v\) is a normalization constant and \(J_1\) is the Bessel function of the first kind. The envelope follows from \(\exp[-\pi^{-1}\int_0^l\tanh(s)\,\mathrm ds]=(\cosh l)^{-1/\pi}\), captures the gap-recovery region, and approaches a constant times \(e^{-l/\pi}\) for \(l\gg1\). For a uniform far-field gap, Gurarie and Radzihovsky obtain the asymptotic form without a weak-coupling expansion~\cite{Gurarie_2007}:

\[
    \binom{U(\mathbf l)}{V(\mathbf l)}
    \propto
    \frac{e^{-l/\pi}}{\sqrt{k_{\rm osc}\xi l}}
    \cos\left(k_{\rm osc}\xi l-\frac{3\pi}{4}\right)
    \binom{e^{i\theta}}{e^{-i\theta}},
    \qquad l\gg1.
\]

Equation~\eqref{eq:approx_cdgm} describes the smooth-core weak-coupling profile, while the second expression gives the uniform-gap far-tail behavior without a weak-coupling expansion. Together, they supply the decay and oscillation scales used below.

An overlap scale follows by assigning the vortex mode \(\ell_{\rm core}\simeq(1+\pi)\xi\), comprising an order-\(\xi\) gap-recovery region and one asymptotic decay length. The chiral edge mode has penetration length \(\ell_{\rm edge}=\pi\xi\)~\cite{Stone_2004}, as derived in Appendix~\ref{sec:no_vortex_disk}. Their sum gives
\begin{equation}
    d\sim\ell_{\rm core}+\ell_{\rm edge}\simeq(2\pi+1)\xi.
\end{equation}
This envelope-based crossover scale lies close to the strongest spectral reorganization around \(d\simeq7\xi\).

The oscillatory tails introduce a possible large-distance continuum modulation. When both asymptotic phase branches contribute, their interference in the squared coupling has the nominal distance period
\begin{equation}
    \lambda_d\simeq\frac{\pi}{k_{\rm osc}}.
\end{equation}
For \(\Delta_0/E_F=0.36\), this gives \(\lambda_d\simeq1.81\xi\). Resolving its period, phase, and amplitude requires finer distance sampling in larger systems, and averaging over edge channels may weaken or eliminate the modulation.

These regional diagnostics show how individual finite-disk eigenstates depart from the polarization pattern of a well-separated core--edge Majorana pair.

\section{Translated core-reference wave packet: spectral reorganization and parity dynamics}
\label{sec:parity_decay}

The preceding section established that the lowest positive-energy eigenstate exhibits nonmonotonic core--edge reorganization as a function of \(d\). To connect this eigenstate behavior to a prescribed core-localized state, we rigidly translate the centered-vortex core Majorana profile to each static vortex position, restrict it to the target disk, and project it onto the retained particle-hole-complete low-energy subspace. Pairing the resulting packet with an ideal stationary remote Majorana defines a fermionic mode whose signed parity memory is expressed through the finite-disk spectral measure.

\begin{figure*}[tb]
\centering
\includegraphics[width=\textwidth]{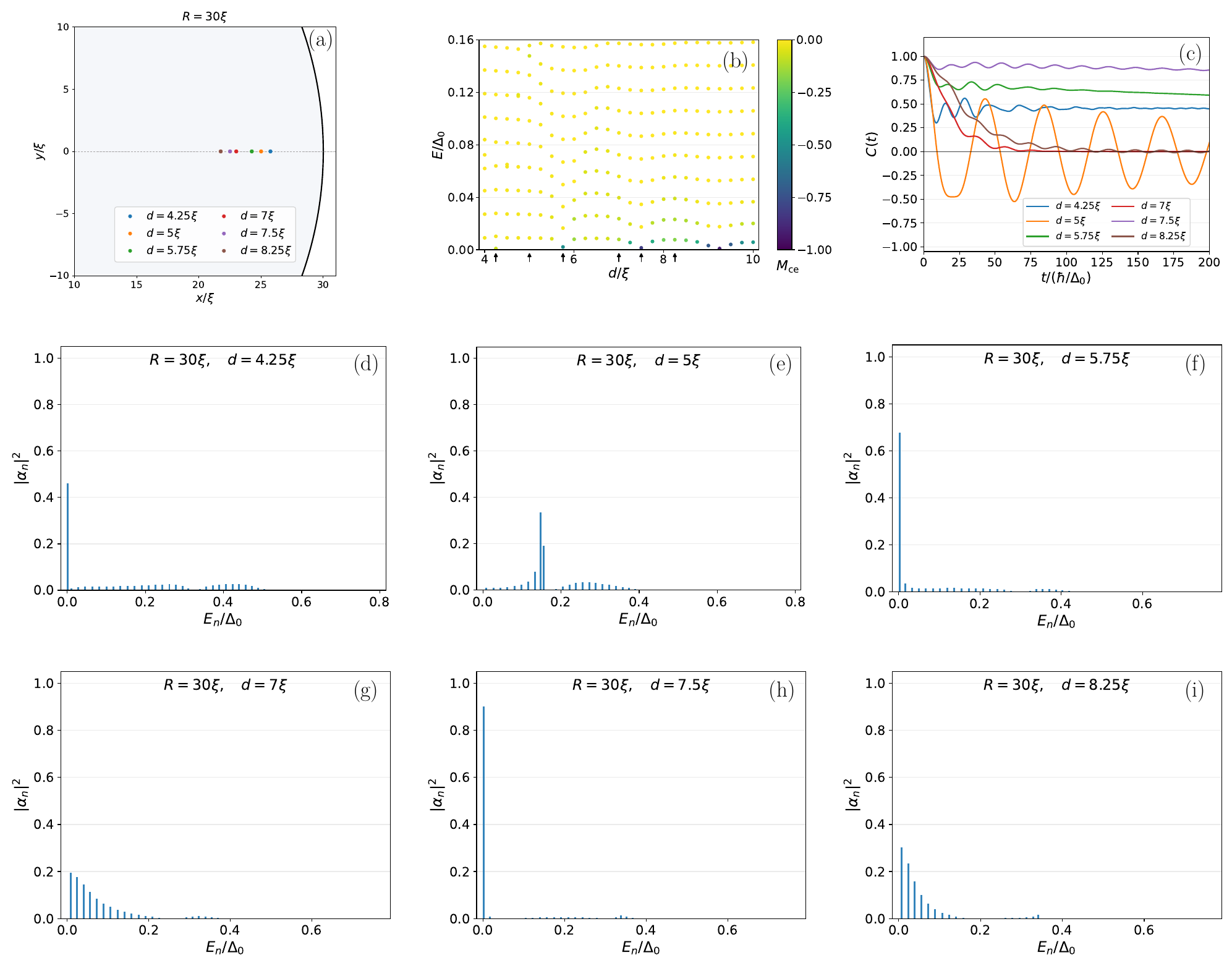}
\caption{Parity dynamics and spectral weights of the translated core-reference packet for \(L=30\), \(\Delta_0/E_F=0.36\), and vortex--boundary separations \(d/\xi=4.25,\ 5,\ 5.75,\ 7,\ 7.5,\ 8.25\). (a) Sampled vortex positions (colored dots); the shading denotes the displayed portion of the superconducting disk. (b) Positive-energy spectra colored by the core--edge Majorana polarization \(M_{\rm ce}\), with arrows marking the six sampled separations. (c) Parity correlator \(C(t)\) for each separation. (d)--(i) Normalized spectral weights \(|\alpha_n|^2\) over the retained \(N=50\) positive-energy eigenstates. At $d/\xi=5.75$ and $7.5$, a dominant near-zero component makes \(C(t)\) vary slowly. At $d/\xi=7$ and $8.25$, fragmentation among several low-energy levels produces rapid coherent dephasing over the displayed interval. At $d/\xi=4.25$, the near-zero component remains the largest despite weak lowest-state core--edge polarization, whereas at $d/\xi=5$, dominant finite-energy weight produces sign-changing oscillations.}
\label{fig:majorana_decay}
\end{figure*}

\subsection{Translated core-reference wave packet}
\label{subsec:bdg_survival}

Let \(\Psi_n=(u_n,v_n)^\top\), \(n=0,\ldots,N-1\), denote the retained \(N=50\) lowest positive-energy eigenstates of \(H_{\rm BdG}\), with energies \(E_n\), and let \(\mathcal C\Psi_n=(v_n^*,u_n^*)^\top\) denote their particle-hole partners. A normalized self-conjugate packet in this particle-hole-complete subspace has the expansion
\begin{equation}
    \phi_{\alpha}
    =
    \frac{1}{\sqrt{2}}\sum_{n=0}^{N-1}
        \left[
            \alpha_n
            \begin{pmatrix}
                u_n\\
                v_n
            \end{pmatrix}
            +\alpha_n^*
            \begin{pmatrix}
                v_n^*\\
                u_n^*
            \end{pmatrix}
        \right],
    \qquad
    \sum_{n=0}^{N-1}\abs{\alpha_n}^2=1.
    \label{eq:truncated_majorana_subspace}
\end{equation}
For the centered vortex, the lowest particle-hole pair generates a two-dimensional real space of self-conjugate combinations. We define the centered core-reference packet as the combination that minimizes \(\langle|\mathbf r-\mathbf r_v|^2\rangle\), obtained by diagonalizing the squared-distance operator \(|\mathbf r-\mathbf r_v|^2\) within the pair subspace. For each static off-center vortex position, we rigidly translate this profile by the vortex displacement, restrict it to the target disk, and project the resulting unnormalized state onto \(\{\Psi_n,\mathcal C\Psi_n\mid n=0,\ldots,N-1\}\) through its overlaps.

Let \(\alpha_{0,n}/\sqrt{2}\) be the overlap of the restricted translated packet with \(\Psi_n\); self-conjugacy fixes its overlap with \(\mathcal C\Psi_n\) as \(\alpha_{0,n}^*/\sqrt{2}\). The norm retained after disk restriction and low-energy projection is
\[
    \mathcal P_N(d)=\left[\sum_{n=0}^{N-1}|\alpha_{0,n}|^2\right]^{1/2}.
\]
Setting \(\alpha_n=\alpha_{0,n}/\mathcal P_N(d)\) gives \(\sum_n|\alpha_n|^2=1\). Over \(4.25\leq d/\xi\leq8.25\), \(\mathcal P_N\geq0.98143\), with the minimum at \(d=4.25\xi\). This retained norm combines boundary clipping and omitted higher-energy weight without separating the two contributions.

\subsection{Parity memory as a spectral survival amplitude}

Let $\psi_n^\dagger$ denote the quasiparticle creation operator associated with $\Psi_n$. The self-conjugate packet in Eq.~\eqref{eq:truncated_majorana_subspace} defines the Majorana operator
\begin{equation}
    \gamma_1=\sum_{n=0}^{N-1}
    \left(\alpha_n\psi_n^\dagger+\alpha_n^*\psi_n\right),
    \qquad \gamma_1^2=1.
    \label{eq:gamma1_decomposition}
\end{equation}
The factor $1/\sqrt{2}$ appearing in the unit-normalized Nambu spinor is absent here because the Majorana convention is $\{\gamma_1,\gamma_1\}=2$. The normalization $\sum_n|\alpha_n|^2=1$ therefore gives $\gamma_1^2=1$.

To define the parity reference, we pair \(\gamma_1\) with an ideal stationary zero-energy Majorana \(\gamma_2\) in a decoupled sample, with \(\{\gamma_i,\gamma_j\}=2\delta_{ij}\). This construction isolates the spectral dynamics of \(\gamma_1\). Such a reference protocol is relevant to measurement-based topological quantum computation, where joint measurements can project pairs of Majorana modes onto definite fermion-parity sectors~\cite{Bonderson_2008,Karzig_2017}.

The fermion $f=(\gamma_1+i\gamma_2)/2$ has parity

\begin{equation}
    P_f=1-2f^\dagger f=-i\gamma_1\gamma_2.
\end{equation}
The signed parity memory is described by the correlator
\begin{equation}
    C(t)\equiv\frac{1}{2}\{P_f(t),P_f(0)\}.
    \label{eq:parity_correlator_def}
\end{equation}
For an initial parity eigenstate satisfying \(P_f\ket{\psi(0)}=\pm\ket{\psi(0)}\), the parity expectation obeys
\begin{equation}
    \begin{aligned}
        \bra{\psi(t)}P_f\ket{\psi(t)}
        &=\bra{\psi(0)}P_f(t)\ket{\psi(0)}\\
        &=\pm\frac{1}{2}\bra{\psi(0)}
        \{P_f(t),P_f(0)\}\ket{\psi(0)}\\
        &=\pm \bra{\psi(0)}C(t)\ket{\psi(0)}.
    \end{aligned}
\end{equation}
Because $\gamma_2$ is stationary and anticommutes with $\gamma_1(t)$, the correlator reduces to a state-independent real number:
\begin{equation}
    \begin{aligned}
        C(t)
        &=-\frac{1}{2}\bigl[
        \gamma_1(t)\gamma_2\gamma_1\gamma_2
        +\gamma_1\gamma_2\gamma_1(t)\gamma_2\bigr]\\
        &=\frac{1}{2}\{\gamma_1(t),\gamma_1\}\\
        &=\sum_{n=0}^{N-1}\abs{\alpha_n}^2\cos(E_n t/\hbar).
    \end{aligned}
    \label{eq:C_to_anticomm}
\end{equation}
For an initial parity eigenstate, \(\langle P_f(t)\rangle=\pm C(t)\), and \(C(t)<0\) corresponds to a coherent inversion of the initial parity expectation.

The finite-disk result can equivalently be written in spectral form. The coefficients in Eq.~\eqref{eq:C_to_anticomm} define the even spectral measure
\begin{equation}
    \rho_N(E)
    =\frac{1}{2}\sum_{n=0}^{N-1}|\alpha_n|^2
    \bigl[\delta(E-E_n)+\delta(E+E_n)\bigr].
    \label{eq:majorana_spectral_measure}
\end{equation}
This measure is normalized, \(\int_{-\infty}^{\infty}\rho_N(E)\,\mathrm dE=1\). Its Fourier transform is precisely the signed parity correlator,
\begin{equation}
    C(t)=\int_{\mathbb R}\rho_N(E)e^{-iEt/\hbar}\,\mathrm dE=\sum_{n=0}^{N-1}|\alpha_n|^2\cos(E_nt/\hbar).
    \label{eq:finite_spectral_survival}
\end{equation}
Equation~\eqref{eq:finite_spectral_survival} identifies $C(t)$ as the particle-hole-symmetrized survival amplitude of the specified packet, with $C(0)=1$. Concentration of spectral weight in one near-zero mode makes \(C(t)\) vary slowly, whereas fragmentation among distinct eigenfrequencies produces coherent dephasing and finite-size recurrences. The relevant dynamics are therefore controlled by the packet's spectral measure, not by its spatial localization alone.

\subsection{Distance-dependent spectral reorganization and parity dynamics}

Figure~\ref{fig:majorana_decay} summarizes the finite-disk results for \(\Delta_0/E_F=0.36\), \(R=30\xi\), and the six sampled vortex--boundary separations. The spectral weights in panels (d)--(i) determine \(C(t)\) in panel (c) through Eq.~\eqref{eq:C_to_anticomm}, while panel (b) shows the corresponding eigenstate polarization. The discrete spectrum produces coherent dephasing and finite-size recurrences. For \(\Delta_0=1\)~meV, the time unit is \(\hbar/\Delta_0\simeq0.66\)~ps.

The spectral response is strongly nonmonotonic in \(d\). At \(d=5.75\xi\) and \(7.5\xi\), most of the packet weight lies in a near-zero state, and \(C(t)\) varies slowly. At \(d=7\xi\) and \(8.25\xi\), the weight is distributed among several low-energy eigenstates, whose relative phases produce rapid coherent dephasing over the displayed interval. Thus, parity memory follows the detailed spectral-weight distribution across the low-energy states.

The two smallest sampled separations show distinct regimes. At \(d=4.25\xi\), the near-zero eigenstate remains the largest spectral component despite its weak regional polarization, supplying a substantial slowly varying contribution to \(C(t)\), while the remaining weights generate faster coherent variation. At \(d=5\xi\), the dominant weight occurs near \(0.1479\Delta_0\), producing sign-changing coherent oscillations.

\section{Discussion and outlook}\label{sec:conclusion}

Our central result is that the retained norm and parity memory provide distinct information about the translated core-reference packet. Across the sampled range, its retained norm remains above \(0.98\), while its spectral measure exhibits near-zero concentration, low-energy fragmentation, or dominant finite-energy weight, depending on \(d\). These distributions produce slow temporal variation, coherent dephasing, and sign-changing parity oscillations, respectively. Thus, the detailed spectral weights control the parity response and generate its nonmonotonic dependence on \(d\).

The eigenstate diagnostics distinguish spatial localization from Majorana separation. For the centered vortex, the lowest state decomposes into complementary core and edge Majorana components, whereas a CdGM state can be strongly core localized without complementary core--edge polarization. The lowest-state energy, regional weight, and polarization vary nonmonotonically with \(d\), with the strongest reorganization among the sampled separations near \(d\simeq7\xi\). This scale is consistent, at the order-of-magnitude level, with the estimated combined core and edge extent \((2\pi+1)\xi\).

For the present \(R=30\xi\) disk, the discrete measure \(\rho_N(E)\) produces unitary dephasing and finite-size recurrences. In the large-system or half-plane limit, the chiral-edge ladder becomes continuous, providing the setting in which vortex--edge coupling may generate a smooth core-localized resonance. The exponential and oscillatory tail structure derived above provides a starting point for analyzing such a resonance.

Natural extensions include a self-consistent treatment of the order parameter to quantify boundary-induced changes in the relative \(p_x\) and \(p_y\) components; finite-size scaling or a direct half-plane calculation to determine whether a smooth resonance emerges and how its width depends on \(d\); explicit adiabatic parallel transport and finite-velocity dynamics to connect the static projection with transport fidelity and nonadiabatic transitions; and disorder, thermal quasiparticles, and phonons to incorporate irreversible decoherence and relaxation.

\section*{Data Availability}

The production code and numerical data supporting the findings of this study are available from the authors upon reasonable request.

\begin{acknowledgments}
This work was supported by the National Key Research and Development Program of China (Grant No.~2022YFA1403403) and the National Natural Science Foundation of China (Grant Nos.~12274441 and 12534004).
\end{acknowledgments}

\appendix

\section{Numerical methods and finite-disk validation}\label{sec:fd_scheme}

This appendix documents the finite-difference implementation and numerical checks for the finite-disk calculations. We first describe the weighted-Hermitian polar-grid discretization, solver diagnostics, and two-grid sensitivity tests in which the radial and angular resolutions are refined simultaneously, then examine sensitivity to the image-amplitude factor and benchmark the edge and centered-vortex limits.

\begin{figure*}[tb]
    \centering
    \includegraphics[width=\textwidth]{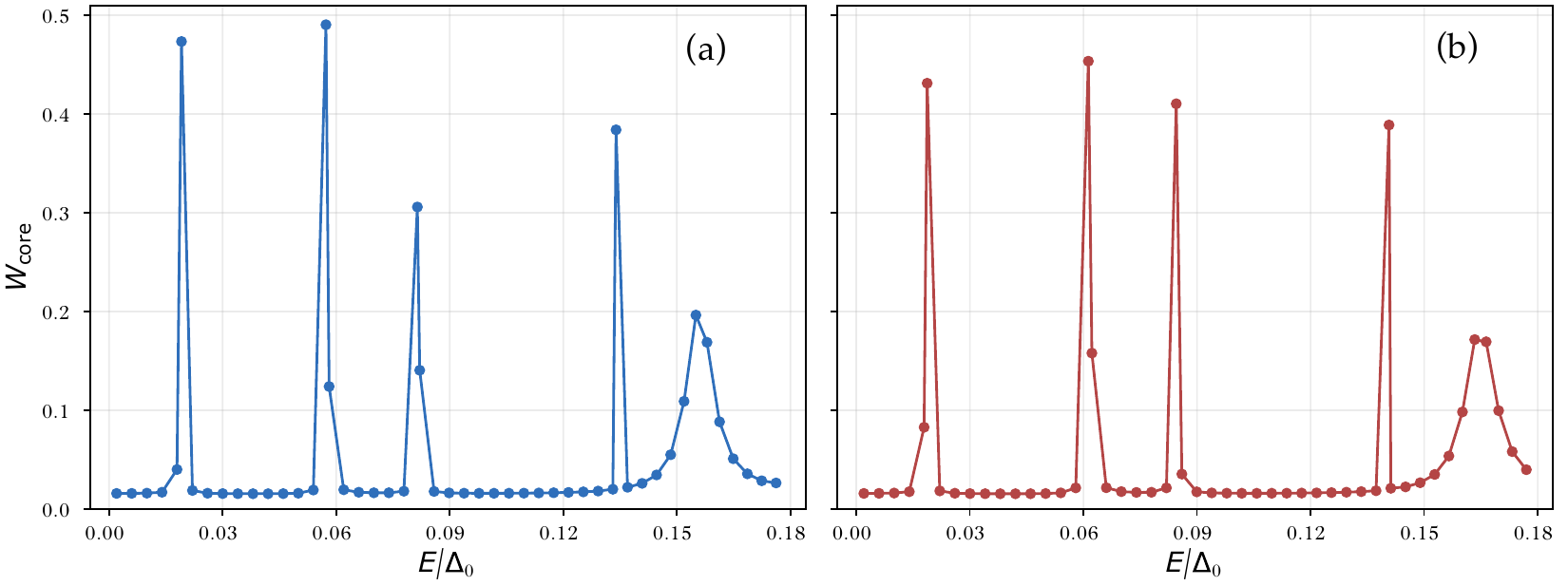}
    \caption{Two-grid sensitivity comparison of the 50 lowest positive-energy states at \(\Delta_0/E_F=0.08\), \(L=30\), and \(d=5\xi\). The vertical coordinate is the normalized core weight \(W_{\rm core}\). Panel (a) on the left uses \((N_l,N_\theta)=(900,900)\), as in the main calculation, and panel (b) on the right uses \((N_l,N_\theta)=(1000,1000)\). The low-core-weight background and the sequence of core-weight peaks appear on both grids, although some peak positions shift by several percent and their weights are redistributed among nearby eigenstates. This test refines \(N_l\) and \(N_\theta\) simultaneously and therefore measures their combined grid sensitivity.}
    \label{fig:grid_refinement_delta008_d5}
\end{figure*}

\begin{figure*}[tb]
    \centering
    \includegraphics[width=\textwidth]{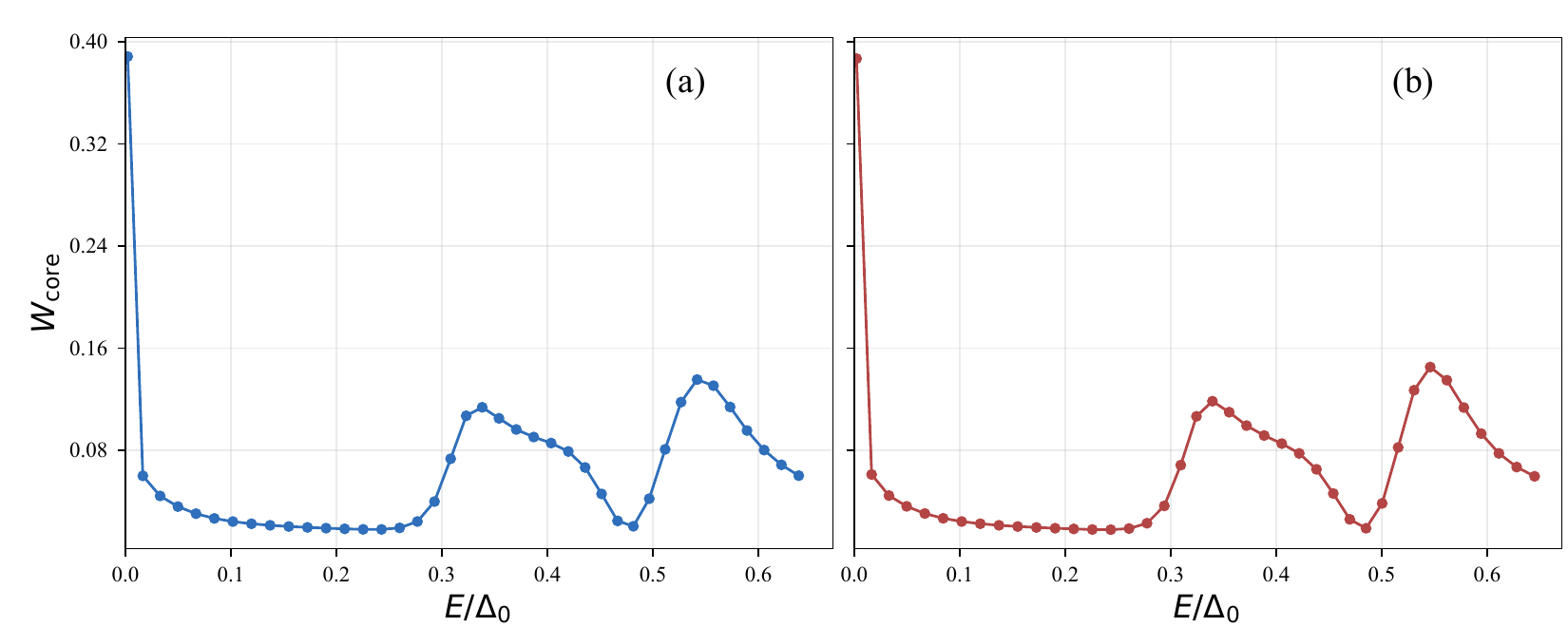}
    \caption{Two-grid sensitivity comparison of the 40 lowest positive-energy states at $\Delta_0/E_F=0.36$, $L=30$, and $d=5.75\xi$. The vertical coordinate is the normalized core weight $W_{\rm core}$. Panel (a) uses $(N_l,N_\theta)=(500,500)$, as in the main calculation, and panel (b) uses $(600,600)$. The near-zero core-weight feature, low-core-weight background, and broader finite-energy structures appear on both grids.}
    \label{fig:grid_refinement_delta036_d575}
\end{figure*}

\begin{figure*}[tb]
    \centering
    \includegraphics[width=\textwidth]{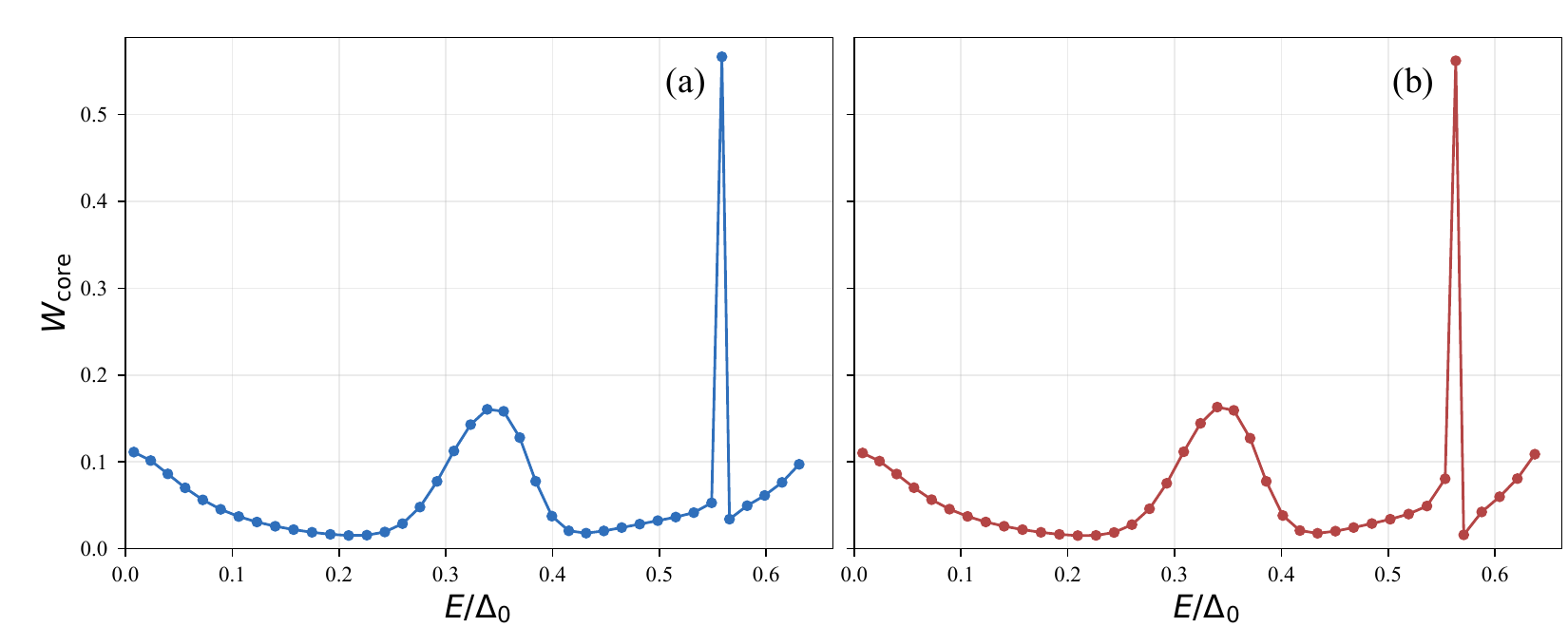}
    \caption{Two-grid sensitivity comparison of the 40 lowest positive-energy states at $\Delta_0/E_F=0.36$, $L=30$, and $d=7\xi$. Panel conventions are as in Fig.~\ref{fig:grid_refinement_delta036_d575}. The low-core-weight background, broad intermediate-energy structure, and sharp high-core-weight peak appear on both grids.}
    \label{fig:grid_refinement_delta036_d7}
\end{figure*}

\begin{figure*}[tb]
    \centering
    \includegraphics[width=\textwidth]{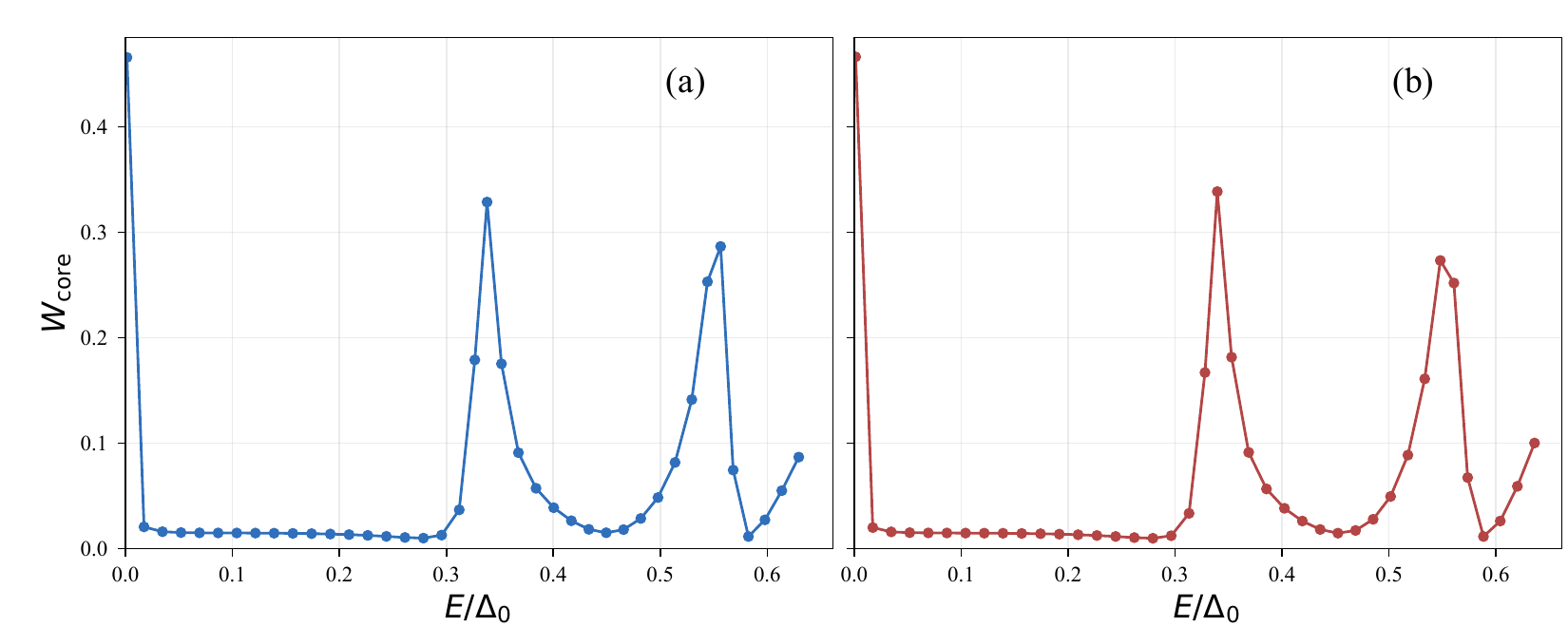}
    \caption{Two-grid sensitivity comparison of the 40 lowest positive-energy states at $\Delta_0/E_F=0.36$, $L=30$, and $d=7.5\xi$. Panel conventions are as in Fig.~\ref{fig:grid_refinement_delta036_d575}. The near-zero core-weight feature and the two principal finite-energy peaks appear on both grids, with only small shifts in their energies and weights.}
    \label{fig:grid_refinement_delta036_d75}
\end{figure*}

\begin{figure*}[tb]
    \centering
    \includegraphics[width=\textwidth]{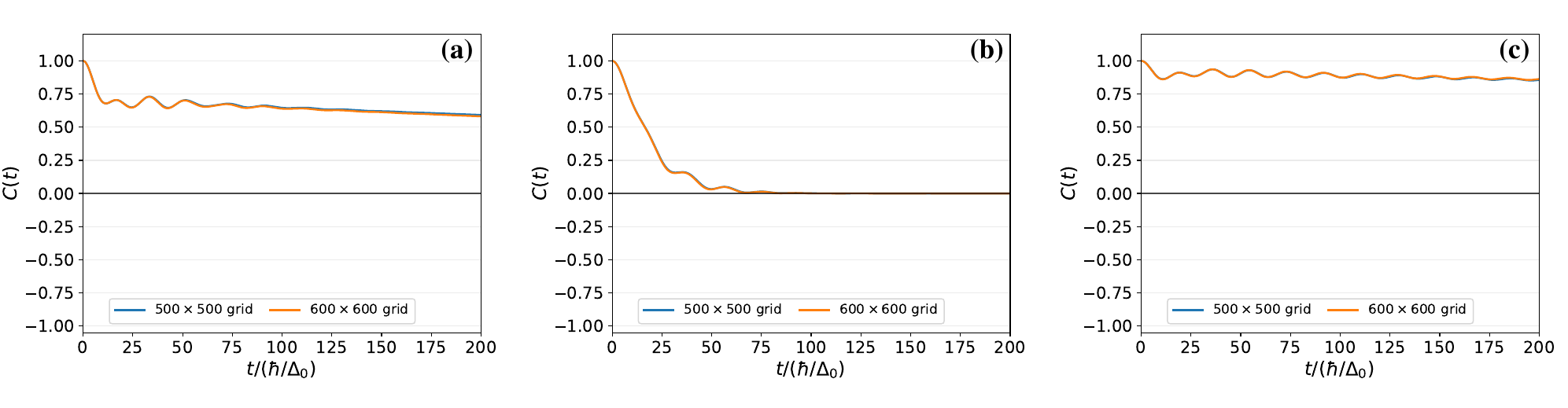}
    \caption{Two-grid sensitivity comparison of the parity correlator \(C(t)\) at \(\Delta_0/E_F=0.36\), \(L=30\), and \(N=50\) retained positive-energy states. Panels (a), (b), and (c) show \(d=5.75\xi\), \(7\xi\), and \(7.5\xi\), respectively. The blue curves use \((N_l,N_\theta)=(500,500)\), as in the main calculation, and the orange curves use \((600,600)\). Over \(0\leq t/(\hbar/\Delta_0)\leq200\), the maximum absolute difference between the two curves remains below \(0.012\) in every panel. As in the spectral comparisons, \(N_l\) and \(N_\theta\) are refined simultaneously.}
    \label{fig:parity_correlator_mesh_refinement}
\end{figure*}

\begin{figure*}[tb]
    \centering
    \includegraphics[width=\textwidth]{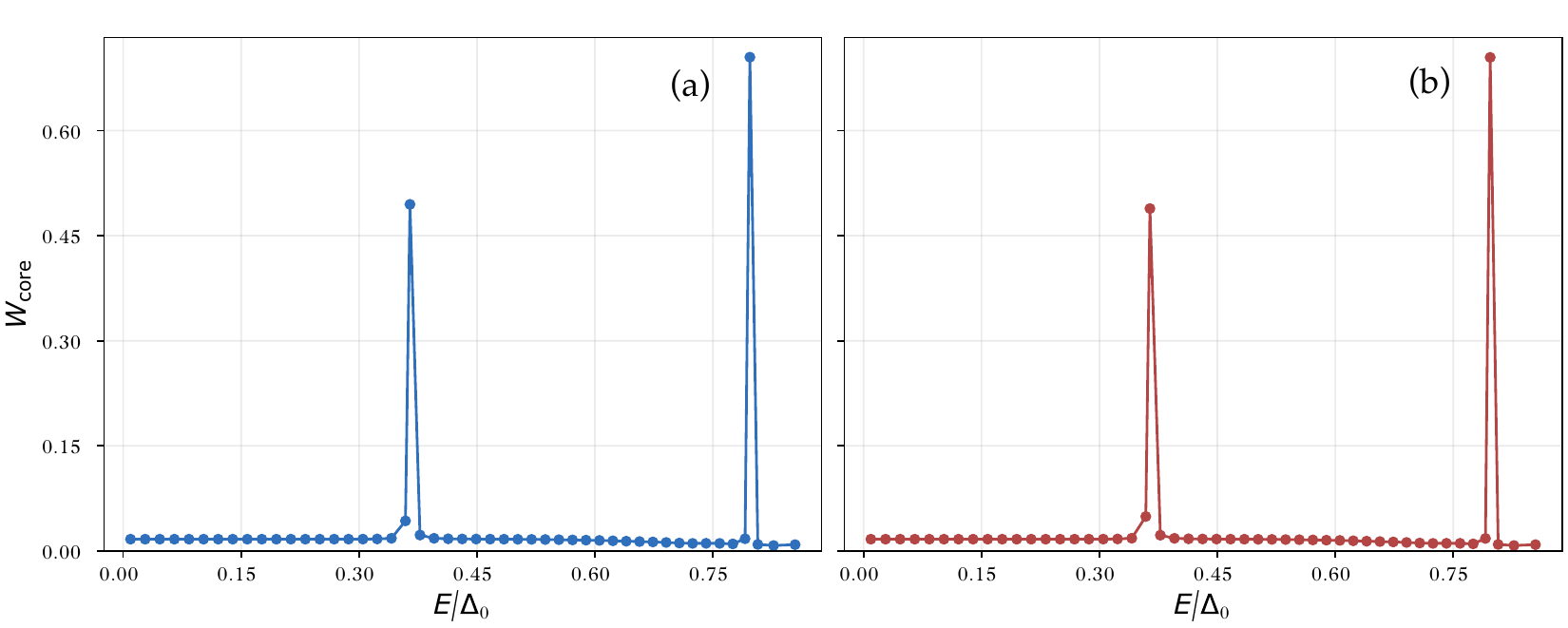}
    \caption{Sensitivity of the positive-energy spectrum to the image-amplitude factor at $\Delta_0/E_F=0.36$, $R=30\xi$, $d=2\xi$, and $(N_l,N_\theta)=(500,500)$. The vertical coordinate is the normalized weight $W_{\rm core}$ in the region closer to the physical vortex than to the disk boundary. Panel (a) on the left uses the full profile in Eq.~\eqref{eq:disk_image}; panel (b) on the right omits only $f(|\mathbf r-\mathbf r_2|/\xi)$ while retaining the image-antivortex phase. Both panels show the 50 lowest positive-energy states. The two prominent core-weight features and low-core-weight background persist under the tested amplitude modification.}
    \label{fig:image_amplitude_sensitivity}
\end{figure*}

\subsection{Polar-grid discretization, solver diagnostics, and two-grid sensitivity}
\label{sec:fd_mesh_refinement}

\begin{table}[tbp]
    \caption{\label{tab:mesh_resolution}Polar-grid resolutions, with dimensionless radial spacing $\Delta l=L/N_l$, angular spacing $\Delta\theta=2\pi/N_\theta$, and Fermi wavelength $\lambda_F=2\pi/k_F$. The last two columns give the numbers of grid points per Fermi wavelength in the radial direction and along the outer boundary, respectively.}
    \begin{ruledtabular}
\begin{tabular}{cccc}
$\Delta_0/E_F$ & $(N_l,N_\theta)$ & $\lambda_F/(\xi\Delta l)$ & $\lambda_F/(R\Delta\theta)$\\
$0.08$ & $(900,900)$ & $23.69$ & $3.77$\\
$0.36$ & $(500,500)$ & $59.22$ & $9.42$\\
\end{tabular}
    \end{ruledtabular}
\end{table}

We discretize Eq.~\eqref{eq:dimensionless_bdg} using radial rings \(l_i=i\Delta l\), \(i=1,\ldots,N_l\), together with one separate point at \(l=0\), and angular points \(\theta_j=j\Delta\theta\), \(j=0,\ldots,N_\theta-1\). Here \(\Delta l=L/N_l\) and \(\Delta\theta=2\pi/N_\theta\). Angular indexing is periodic, and the outermost ring satisfies the hard-wall condition \(U_{N_l,j}=V_{N_l,j}=0\). At points away from the origin, the polar differential operators
\begin{equation}
    \begin{aligned}
        \partial_{\pm}
        &=e^{\pm i\theta}
        \left(\partial_l\pm\frac{i}{l}\partial_\theta\right),\\
        \nabla^2
        &=\partial_l^2+\frac{1}{l}\partial_l
        +\frac{1}{l^2}\partial_\theta^2.
    \end{aligned}
\end{equation}
are approximated by centered second-order differences. For a grid function \(F_{i,j}\),
\begin{equation}
    \begin{aligned}
        (\partial_lF)_{i,j}
        &=\frac{F_{i+1,j}-F_{i-1,j}}{2\Delta l},\\
        (\partial_\theta F)_{i,j}
        &=\frac{F_{i,j+1}-F_{i,j-1}}{2\Delta\theta},\\
        (\partial_l^2F)_{i,j}
        &=\frac{F_{i+1,j}-2F_{i,j}+F_{i-1,j}}{(\Delta l)^2},\\
        (\partial_\theta^2F)_{i,j}
        &=\frac{F_{i,j+1}-2F_{i,j}+F_{i,j-1}}{(\Delta\theta)^2}.
    \end{aligned}
\end{equation}

The coordinate singularity requires separate origin stencils. Denoting the grid-function value at the origin by $F_0$, a Cartesian Taylor expansion and angular average over the first ring give
\begin{equation}
    \left.\nabla^2F\right|_0
    =\frac{4}{(\Delta l)^2}
    \left(\frac{1}{N_\theta}\sum_{j=0}^{N_\theta-1}F_{1,j}-F_0\right)
    +\mathcal O\bigl((\Delta l)^2\bigr),
    \label{eq:lap_pole}
\end{equation}
while angularly weighted ring averages give
\begin{equation}
    \left.\partial_\pm F\right|_0
    =\frac{2}{N_\theta\Delta l}
    \sum_{j=0}^{N_\theta-1}F_{1,j}e^{\pm i\theta_j}
    +\mathcal O\bigl((\Delta l)^2\bigr).
    \label{eq:dplus_pole}
\end{equation}
Both origin stencils have \(O((\Delta l)^2)\) local truncation error for smooth fields, consistent with the centered second-order scheme away from the origin.

Let \(W\) be the diagonal spatial quadrature matrix with origin weight \(w_0=\pi(\Delta l/2)^2\) and ring weights \(w_{i,j}=l_i\Delta l\Delta\theta\), and define the Nambu-space weight \(\mathcal W=\operatorname{diag}(W,W)\). Writing the discrete BdG operator as \(H_{\rm FD}=\left(\begin{smallmatrix}K_{\rm FD}&P_{\rm FD}\\-P_{\rm FD}^*&-K_{\rm FD}\end{smallmatrix}\right)\), the normal block satisfies \(WK_{\rm FD}=K_{\rm FD}^\dagger W\). Weighted antisymmetrization of the pairing block yields the Hermitian representation
\begin{equation}
    \begin{aligned}
        \widetilde H&=
        \begin{pmatrix}
            A&B\\
            B^\dagger&-A
        \end{pmatrix},\\
        A&=W^{1/2}K_{\rm FD}W^{-1/2},\\
        B&=\frac{1}{2}W^{1/2}P_{\rm FD}W^{-1/2}
        -\frac{1}{2}W^{-1/2}P_{\rm FD}^{\mathsf T}W^{1/2}.
    \end{aligned}
    \label{eq:weighted_fd_operator}
\end{equation}
Eigenvectors of \(\widetilde H\) are transformed back to the physical grid with \(\operatorname{diag}(W^{-1/2},W^{-1/2})\). The block construction is Hermitian and particle-hole symmetric: the weighted upper-right block is antisymmetric, and the lower-left block is its Hermitian conjugate, equivalently the negative complex conjugate required by BdG particle-hole symmetry. In the implementation, \(\widetilde H\) is additionally averaged with \(\widetilde H^\dagger\) before diagonalization to remove roundoff-level Hermiticity defects. The mesh resolutions used in the main calculations are summarized in Table~\ref{tab:mesh_resolution}.

We solve \(\widetilde H\) using the sparse shift-invert routine \texttt{eigsh}, with target shift \(\sigma=0\) and requested tolerance \(10^{-8}\), and retain the 50 positive eigenvalues closest to zero. Each returned transformed eigenpair \((\epsilon_n,q_n)\) must satisfy \(\lVert\widetilde Hq_n-\epsilon_nq_n\rVert_2/\max(1,|\epsilon_n|)<10^{-8}\). After transformation back to the physical grid, the spinors \(\Psi_n\) must satisfy weighted normalization and orthogonality: for \(G_{mn}=\Psi_m^\dagger\mathcal W\Psi_n\), we require \(\max_n|G_{nn}-1|<10^{-8}\) and \(\max_{m\ne n}|G_{mn}|<10^{-8}\). We define the weighted-Hermiticity defect of the full BdG operator as \(\max_{ij}|(H_{\rm FD}^\dagger\mathcal W-\mathcal W H_{\rm FD})_{ij}|\). The validation records for \(\Delta_0/E_F=0.08\) and \(0.36\) give maximum defects of \(4.4\times10^{-16}\) and \(2.7\times10^{-15}\), respectively.

The \(\Delta_0/E_F=0.08\) calculation has the coarsest outer-boundary resolution in Fermi-wavelength units and is therefore the most demanding discretization. Figure~\ref{fig:grid_refinement_delta008_d5} compares \(W_{\rm core}(E)\) at \(L=30\) and \(d=5\xi\) on \((N_l,N_\theta)=(900,900)\) and \((1000,1000)\) grids. The low-core-weight edge ladder and sequence of pronounced core-weight features appear on both grids. The first four prominent peaks shift from \(E/\Delta_0=0.01915\), \(0.05731\), \(0.08146\), and \(0.13394\) to \(0.01872\), \(0.06130\), \(0.08452\), and \(0.14072\), respectively, quantifying the grid sensitivity of individual peak positions and weights at \(d=5\xi\).

For \(\Delta_0/E_F=0.36\), Figs.~\ref{fig:grid_refinement_delta036_d575}--\ref{fig:grid_refinement_delta036_d75} compare the 40 lowest positive-energy states obtained on \((500,500)\) and \((600,600)\) grids at \(d=5.75\xi\), \(7\xi\), and \(7.5\xi\). At all three separations, the low-core-weight background and prominent core-weight features appear in both spectra, and the two-grid spectra remain close over the displayed energy window. These two-grid comparisons quantify the sensitivity of \(W_{\rm core}(E)\) to the simultaneous change in radial and angular resolution at the sampled separations in the rapid-reorganization region.

Figure~\ref{fig:parity_correlator_mesh_refinement} compares the translated-packet parity correlator on the same \(500^2\) and \(600^2\) grids. Over \(0\leq t/(\hbar/\Delta_0)\leq200\), the maximum absolute differences between the two \(C(t)\) curves are \(0.0113\), \(0.0069\), and \(0.0073\) at \(d=5.75\xi\), \(7\xi\), and \(7.5\xi\), respectively. These values quantify the sensitivity of \(C(t)\) to the simultaneous change in radial and angular resolution over the tested separations and time window.

\subsection{Sensitivity to the image-amplitude factor}\label{app:image_amplitude_sensitivity}

The image phase in Eq.~\eqref{eq:disk_image} enforces the condensate no-normal-current boundary condition, while the image factor \(f(|\mathbf r-\mathbf r_2|/\xi)\) belongs to the prescribed amplitude ansatz. We test its influence at \(d=2\xi\), where the image-amplitude factor deviates more strongly from unity than at the six translated-packet separations \(d\geq4.25\xi\). The calculation is repeated with this factor set to unity while retaining the physical-vortex amplitude, image-antivortex phase, and all other parameters. Figure~\ref{fig:image_amplitude_sensitivity} shows that the two largest-core-weight features shift only slightly and retain comparable weights, while the low-core-weight background persists. The energy-resolved core-weight pattern therefore remains qualitatively unchanged under this amplitude modification.

\subsection{Uniform-disk edge benchmark}\label{sec:no_vortex_disk}

For a uniform disk without a vortex, rotational symmetry permits $U=e^{im\theta}U_m(l)$ and $V=e^{i(m-1)\theta}V_m(l)$ for $m\in\mathbb Z$, with half-integer edge momentum $j=m-\tfrac12$. The large-$L$ boundary-layer solution is
\begin{equation}
    \begin{aligned}
        \binom{U}{V}&\simeq\frac{\mathcal N e^{-(L-l)/\pi}}{\sqrt l}\sin[k_{\rm osc}\xi(L-l)]\binom{e^{i(j+1/2)\theta}}{e^{i(j-1/2)\theta}},\\
        \frac{E_j}{\Delta_0}
        &=-\frac{\pi\Delta_0}{2\mu}\frac{j}{L}.
    \end{aligned}
    \label{eq:edge_wavefunction_asymptotic}
\end{equation}
Here \(\mathcal N\) is a normalization constant. Equation~\eqref{eq:edge_wavefunction_asymptotic} uses the \(q=0\) decay and oscillation scales. Let \(x=R-r\geq0\) denote the inward normal distance and take the normal dependence to be \(e^{ik_\perp x}\). For a locally flat edge mode with \(q=j/R\), the decaying branches
\[
    k_\perp(q)
    =\pm\sqrt{k_{\rm osc}^2-q^2}+\frac{i}{\pi\xi}
\]
follow from
\[
    \left[
        \frac{\hbar^2(k_\perp^2+q^2)}{2m_e}-\mu
    \right]^2
    +\frac{\Delta_0^2}{k_F^2}(k_\perp^2+q^2)
    =\frac{\Delta_0^2}{k_F^2}q^2.
\]
The two branches combine to give the sine factor in Eq.~\eqref{eq:edge_wavefunction_asymptotic}. Disk-curvature corrections to the boundary-layer result enter at \(O(L^{-2})\) for fixed \(j\), where \(L=R/\xi\). The dispersion gives adjacent-level spacing \(\delta E=\pi\Delta_0^2/(2\mu L)\). We benchmark these large-\(L\) predictions using an independent Bessel-basis projection assembled with standard recurrence and Lommel identities~\cite{watson1995bessel,NIST:DLMF}. At \(\Delta_0/E_F=0.36\) and \(L=30\), the Bessel-basis calculation gives the lowest absolute energy \(9.455281\times10^{-3}\Delta_0\), compared with the asymptotic value \(9.424778\times10^{-3}\Delta_0\). The difference is \(0.32\%\); the benchmark spectrum contains the expected chiral in-gap branch.

\subsection{Centered-vortex benchmark}\label{sec:centered_vortex}

\begin{table}[tbp]
    \caption{\label{tab:energy_benchmarks}Reference and numerical edge and core energy scales at $\Delta_0/E_F=0.36$ and $L=30$.}
    \begin{ruledtabular}
\begin{tabular}{lccc}
Observable & Reference & Numerical & Difference\\
Edge $|E|/\Delta_0$ & $0.0094248$ & $0.0094553$ & $0.32\%$\\
Edge $\delta E/\Delta_0$ & $0.0188496$ & $0.0189106$ & $0.32\%$\\
Core $E/\Delta_0$ & $0.3379$ & $0.3366$ & $0.38\%$\\
\end{tabular}
    \end{ruledtabular}
\end{table}

For a centered unit-winding vortex, the vortex phase winding shifts the rotationally symmetric angular ansatz to $U=e^{i(m+1)\theta}U_m(l)$ and $V=e^{i(m-1)\theta}V_m(l)$. The radial problem can be projected independently onto a Bessel basis that is regular at the origin and vanishes at $l=L$~\cite{Hayashi_1998,Gygi_1991}. For $f(l)=\tanh l$, the weak-coupling smooth-core approximation to the isolated-vortex Majorana is~\cite{Cheng_2009}
\begin{equation}
    U_0(l)\simeq V_0(l)\simeq
    \mathcal N_0(\cosh l)^{-1/\pi}J_1(k_F\xi l),
    \label{eq:vortex_core_zero_mode}
\end{equation}
where $\mathcal N_0$ is a normalization constant. For a uniform far-field gap, the solution without the weak-coupling expansion instead oscillates with $k_{\rm osc}$ and has the asymptotic form $e^{-l/\pi}\cos(k_{\rm osc}\xi l-3\pi/4)/\sqrt{k_{\rm osc}\xi l}$~\cite{Gurarie_2007}. Defining $\mathcal K(l)=\pi^{-1}\int_0^l f(s)\,\mathrm ds$, the CdGM ladder has spacing~\cite{Caroli_1964,Kopnin_1991}
\begin{equation}
    \begin{aligned}
        E_m^{\rm core}&\simeq m\omega_v,\\
        \omega_v&\simeq\frac{\Delta_0}{k_F\xi}\frac{\int_0^\infty[f(l)/l]e^{-2\mathcal K(l)}\,\mathrm dl}{\int_0^\infty e^{-2\mathcal K(l)}\,\mathrm dl}\simeq1.02\frac{\Delta_0^2}{\mu}.
    \end{aligned}
    \label{eq:vortex_core_spacing}
\end{equation}
At the moderately coupled value \(\Delta_0/\mu=0.36\), the asymptotic expression gives \(\omega_v=0.3674\Delta_0\), while the radial Bessel-basis calculation gives the first positive core excitation at \(0.3379\Delta_0\). The asymptotic value sets the expected core-energy scale, and Table~\ref{tab:energy_benchmarks} uses the Bessel-basis value as the independent reference for the polar-grid result.
The first two rows compare the large-\(L\) edge theory with the Bessel-basis calculation, while the last compares the centered-vortex Bessel-basis and polar-grid calculations. These independent comparisons support the accuracy of the edge energy scale, edge-level spacing, and centered-vortex core excitation used in the finite-disk analysis at the tested parameters.

\section{Regional Majorana polarization and partition robustness}\label{app:majorana_polarization}

This appendix derives the regional Majorana-polarization identities used in Sec.~\ref{sec:spectrum} and clarifies their basis-independent interpretation. It then supplements the distance-dependent diagnostics with representative eigenstate densities and tests their sensitivity to the core--edge partition.

\begin{figure*}[tb]
    \centering
    \includegraphics[width=\textwidth]{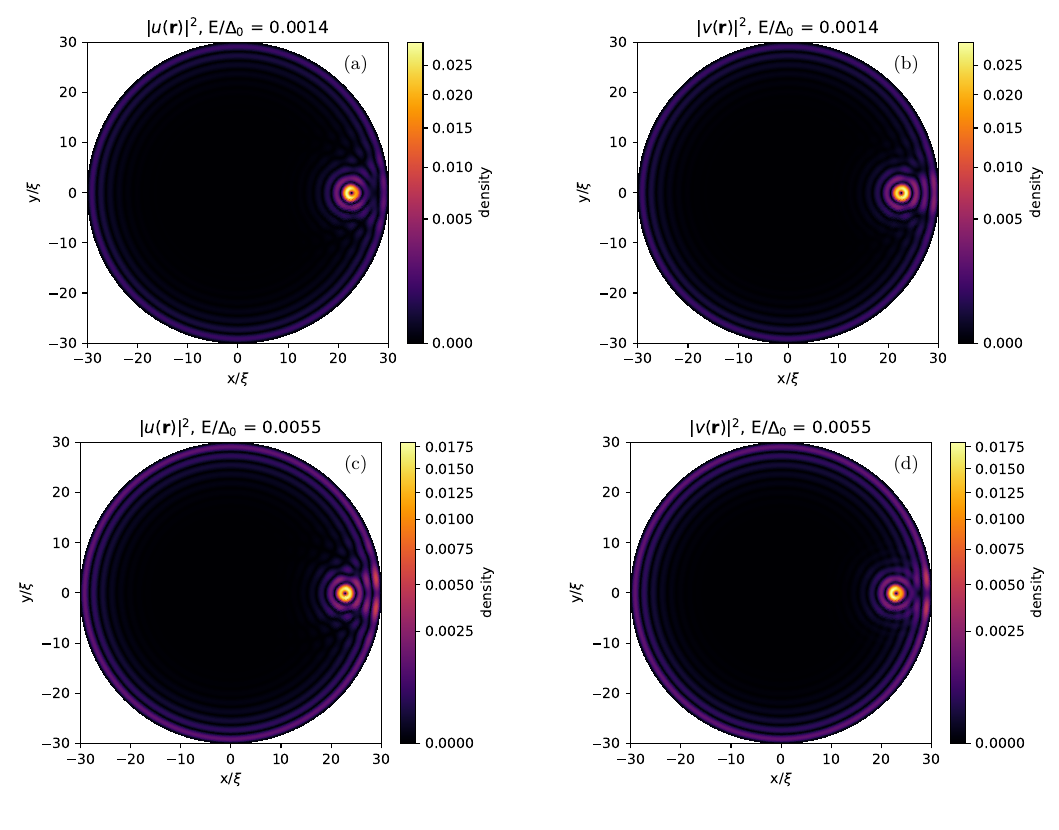}
    \caption{Lowest positive-energy eigenstate near the region of rapid core--edge reorganization for $\Delta_0/E_F=0.36$ and $R=30\xi$. (a) Particle density $|u|^2$ and (b) hole density $|v|^2$ at $d=7.5\xi$, where $E_0/\Delta_0=0.00141$, $M_{\rm ce}=-0.8159$, and $W_{\rm core}=0.4657$. (c) $|u|^2$ and (d) $|v|^2$ at $d=7.25\xi$, where $E_0/\Delta_0=0.00549$, $M_{\rm ce}=-0.3882$, and $W_{\rm core}=0.2904$. Color scales are shared within each row but differ between the two separations.}
    \label{fig:fd_polar_d7p5_state0}
\end{figure*}

\subsection{General polarization identities}

For a normalized positive-energy BdG eigenstate \(\psi=[u,v]^\top\) with energy $E$, particle-hole symmetry supplies the partner \(\mathcal C\psi=[v^*,u^*]^\top\) with energy \(-E\). Because \(E\neq0\), Hermiticity ensures that these states are orthogonal. Their two-dimensional subspace admits the self-conjugate orthonormal basis
\begin{equation}
    \phi_1=\frac{1}{\sqrt{2}}
    \begin{pmatrix}
        u+v^*\\
        v+u^*
    \end{pmatrix},
    \qquad
    \phi_2=\frac{i}{\sqrt{2}}
    \begin{pmatrix}
        u-v^*\\
        v-u^*
    \end{pmatrix}.
\end{equation}
Their time evolution is
\begin{equation}
    \begin{pmatrix}
        \phi_1(t)&\phi_2(t)
    \end{pmatrix}
    =
    \begin{pmatrix}
        \phi_1&\phi_2
    \end{pmatrix}
    \begin{pmatrix}
        \cos(Et/\hbar)&\sin(Et/\hbar)\\
        -\sin(Et/\hbar)&\cos(Et/\hbar)
    \end{pmatrix}.
\end{equation}
Thus, every nonzero-energy fermionic mode decomposes into two self-conjugate components that rotate into one another at angular frequency \(E/\hbar\); identifying spatially separated Majorana components, even near zero energy, additionally requires a spatial diagnostic.

For a complementary partition \(\Omega\cup\bar\Omega\) of the disk with \(W_\Omega>0\) and \(W_{\bar\Omega}>0\), define the probability weight \(W_A\), polarization numerator \(n_A\), and normalized regional Majorana polarization \(P_A\), for \(A\in\{\Omega,\bar\Omega\}\), by
\begin{equation}
    \begin{aligned}
        W_A&=\int_A\! \mathrm d^2r\,(|u|^2+|v|^2),\\
        n_A&=2\int_A\! \mathrm d^2r\,(uv)^*,\\
        P_A&=\frac{n_A}{W_A},
    \end{aligned}
    \label{eq:generic_regional_majorana_polarization}
\end{equation}
Normalization gives \(W_\Omega+W_{\bar\Omega}=1\), while \(2|uv|\leq|u|^2+|v|^2\) gives \(|P_A|\leq1\). The phase-convention-independent combination is
\begin{equation}
    M_{\Omega:\bar\Omega}
    =\operatorname{Re} \bigl(P_\Omega P_{\bar\Omega}^*\bigr).
    \label{eq:generic_nonlocal_majorana_polarization}
\end{equation}
Under the global eigenstate rephasing $\psi\rightarrow e^{i\chi}\psi$, both regional polarizations acquire the same phase $e^{-2i\chi}$, which cancels in $M_{\Omega:\bar\Omega}$.

Orthogonality of the states at $\pm E$ implies
\begin{equation}
    0=2\int\! \mathrm d^2r\,(uv)^*
    =n_\Omega+n_{\bar\Omega},
    \qquad
    n_{\bar\Omega}=-n_\Omega,
\end{equation}
and hence
\begin{equation}
    M_{\Omega:\bar\Omega}
    =-\frac{|n_\Omega|^2}{W_\Omega W_{\bar\Omega}},
    \qquad
    -1\leq M_{\Omega:\bar\Omega}\leq0.
    \label{eq:generic_majorana_polarization_bound}
\end{equation}
When \(n_\Omega\neq0\), the two regional polarizations differ in phase by \(\pi\); in all cases, \(M_{\Omega:\bar\Omega}\leq0\). The bound \(|n_\Omega|\leq\min(W_\Omega,W_{\bar\Omega})\) further gives \(M_{\Omega:\bar\Omega}\geq-1\). Thus the diagnostic measures complementary Majorana separation rather than localization in either region alone.

A basis-independent interpretation follows by restricting the two-dimensional self-conjugate subspace to \(\Omega\). Define its regional overlap matrix
\begin{equation}
    (G_\Omega)_{ij}
    =\int_\Omega\! \mathrm d^2r\,
    \phi_i^\dagger(\mathbf r)\phi_j(\mathbf r),
    \qquad i,j=1,2.
\end{equation}
Under \(\psi\to e^{i\chi}\psi\), the basis \((\phi_1,\phi_2)\) undergoes a real orthogonal rotation, so the eigenvalues \(\lambda_+\geq\lambda_-\) of \(G_\Omega\) are invariant. For a normalized real recombination of \(\phi_1\) and \(\phi_2\), the regional weight is the corresponding quadratic form of \(G_\Omega\); hence \(\lambda_+\) and \(\lambda_-\) are the largest and smallest weights attainable in \(\Omega\). In the basis that diagonalizes \(G_\Omega\),
\begin{equation}
    W_\Omega=\frac{\lambda_++\lambda_-}{2},
    \qquad
    |n_\Omega|=\frac{\lambda_+-\lambda_-}{2},
\end{equation}
which gives
\begin{equation}
    M_{\Omega:\bar\Omega}
    =-\frac{(\lambda_+-\lambda_-)^2}
    {(\lambda_++\lambda_-)(2-\lambda_+-\lambda_-)}.
    \label{eq:generic_majorana_polarization_overlap}
\end{equation}
Thus, \(M_{\Omega:\bar\Omega}\simeq-1\) when \(\lambda_+\simeq1\) and \(\lambda_-\simeq0\), corresponding to one self-conjugate recombination concentrated in \(\Omega\) and its orthogonal partner concentrated in \(\bar\Omega\). Conversely, \(\lambda_+\simeq\lambda_-\) gives \(M_{\Omega:\bar\Omega}\simeq0\), encompassing both same-region localization and generic extended states.

\subsection{Eigenstates in the rapid-reorganization regime}

Figure~\ref{fig:fd_polar_d7p5_state0} compares the lowest-state particle and hole densities at \(d=7.5\xi\) and \(7.25\xi\), two neighboring sampled separations in the rapid-reorganization region identified in Fig.~\ref{fig:energy_vs_d_observables}. Although the values of \(d\) differ by \(0.25\xi\), the eigenstates have substantially different density distributions between the vortex core and boundary.

\subsection{Partition robustness}

\begin{table}[tbp]
    \caption{\label{tab:partition_robustness}Core--edge Majorana polarization of the lowest positive-energy state for several regional partitions. The superscript $(a{:}b)$ denotes the boundary $|\mathbf r-\mathbf r_v|/[R-|\mathbf r|]=a/b$.}
    \begin{ruledtabular}
\begin{tabular}{cccccc}
$d/\xi$ & $M_{\rm ce}^{(1{:}3)}$ & $M_{\rm ce}^{(1{:}2)}$ & $M_{\rm ce}^{(1{:}1)}$ & $M_{\rm ce}^{(2{:}1)}$ & $M_{\rm ce}^{(3{:}1)}$\\
$30$  & $-0.9823$ & $-0.9951$ & $-0.9997$ & $-0.9964$ & $-0.9806$\\
$10$  & $-0.3667$ & $-0.4297$ & $-0.4612$ & $-0.4659$ & $-0.4602$\\
$7.5$ & $-0.5643$ & $-0.6798$ & $-0.8159$ & $-0.8265$ & $-0.7998$\\
$5$   & $-0.0055$ & $-0.0095$ & $-0.0193$ & $-0.0394$ & $-0.0550$\\
$2$   & $-0.0015$ & $-0.0044$ & $-0.0169$ & $-0.0379$ & $-0.0582$\\
\end{tabular}
    \end{ruledtabular}
\end{table}

To assess sensitivity to the regional boundary, we generalize Eq.~\eqref{eq:core_edge_partition} by defining \(\Omega_{\rm core}^{(a:b)}\) through \(|\mathbf r-\mathbf r_v|/[R-|\mathbf r|]<a/b\), with \(\Omega_{\rm edge}^{(a:b)}\) its complement. Table~\ref{tab:partition_robustness} reports the lowest-state polarization for five such partitions.
Across all tested partitions, the centered-vortex value remains close to \(-1\), while the values at \(d=5\xi\) and \(2\xi\) remain close to zero. The values at \(d=10\xi\) and \(7.5\xi\) show greater sensitivity to the regional boundary. Thus, changing the partition affects the intermediate magnitudes while preserving the endpoint contrast between a well-separated core--edge pair at large \(d\) and weak complementary separation near the boundary.

\bibliography{ref}

\end{document}